\documentclass[dvipsnames]{lmcs} 
\pdfoutput=1

\usepackage{lastpage}
\lmcsdoi{16}{1}{10}
\lmcsheading{}{\pageref{LastPage}}{}{}%
{Feb.~06,~2019}{Feb.~13,~2020}{}

\keywords{homotopy type theory, higher inductive types, inductive types}

\usepackage{microtype}
\usepackage{proof}
\usepackage{amsmath}
\usepackage{amssymb}
\usepackage{stmaryrd}
\usepackage{hyperref}
\usepackage{url}

\makeatletter
\DeclareRobustCommand{\sqcdot}{\mathbin{\mathpalette\morphic@sqcdot\relax}}
\newcommand{\morphic@sqcdot}[2]{%
  \sbox\z@{$\m@th#1\centerdot$}%
  \ht\z@=.33333\ht\z@
  \vcenter{\box\z@}%
}
\makeatother

\newcommand{\U}{\mathsf{U}}
\newcommand{\El}{\mathsf{El}}

\newcommand{\ra}{\rightarrow}

\newcommand{\Set}{\mathsf{Set}}

\newcommand{\Con}{\mathsf{Con}}
\newcommand{\Ty}{\mathsf{Ty}}
\newcommand{\Tm}{\mathsf{Tm}}
\newcommand{\Sub}{\mathsf{Sub}}

\newcommand{\id}{\mathsf{id}}

\newcommand{\Nat}{\mathsf{Nat}}
\newcommand{\blank}{\mathord{\hspace{1pt}\text{--}\hspace{1pt}}}

\newcommand{\A}{\mathsf{A}}
\newcommand{\M}{\mathsf{M}}
\newcommand{\D}{\mathsf{D}}
\renewcommand{\S}{\mathsf{S}}

\newcommand{\proj}{\mathsf{proj}}

\newcommand{\refl}{\mathsf{refl}}

\newcommand{\zero}{\mathsf{zero}}
\newcommand{\suc}{\mathsf{suc}}

\newcommand{\Elim}{\mathsf{Elim}}

\newcommand{\coe}{\mathsf{coe}}

 \newcommand{\Bool}{\mathsf{Bool}}

 \newcommand{\J}{\mathsf{J}}

 \newcommand{\E}{\mathsf{S}}

\newcommand{\inv}{\mathsf{inv}}
\newcommand{\con}{\mathsf{con}}

\newcommand{\Int}{\mathsf{Int}}
\renewcommand{\in}{\mathbin{\hat:}}
\renewcommand{\hat}[1]{{\color{BrickRed}{#1}}}
\newcommand{\blc}[1]{{\color{Black}{#1}}}
\newcommand{\vdashh}{\mathbin{\hat\vdash}}
\newcommand{\rah}{\mathbin{\hat\ra}}
\newcommand{\commah}{\hat,\,}
\newcommand{\timesh}{\mathbin{\hat\times}}
\newcommand{\eqh}{\mathbin{\hat=}}
\newcommand{\TR}{\hat{\mathsf{tr}}}
\newcommand{\ap}{\hat{\mathsf{ap}}}
\newcommand{\apd}{\hat{\mathsf{apd}}}
\renewcommand{\tt}{\hat{\mathsf{tt}}}

\newcommand{\Type}{\hat{\mathsf{Type}}}

\newcommand{\semicol}{\hat;\,}
\newcommand{\targetass}{\hat{\Gamma}\semicol}
\newcommand{\Alg}{\mathsf{Alg}}
\newcommand{\DisplayedAlg}{\mathsf{DisplayedAlg}}
\newcommand{\Morphism}{\mathsf{Morphism}}
\newcommand{\Section}{\mathsf{Section}}

\renewcommand{\inv}{\mathsf{inv}}

\theoremstyle{plain} 

\begin{document}

\title{Signatures and Induction Principles for Higher Inductive-Inductive Types}

\author[A.~Kaposi]{Ambrus Kaposi}	
\address{Department of Programming Languages and Compilers, E{\"o}tv{\"o}s Lor{\'a}nd University, Budapest, Hungary}	
\email{akaposi@inf.elte.hu}  

\author[A.~Kov{\'a}cs]{Andr{\'a}s Kov{\'a}cs}	
\address{Department of Programming Languages and Compilers, E{\"o}tv{\"o}s Lor{\'a}nd University, Budapest, Hungary}	
\email{kovacsandras@inf.elte.hu}  
\thanks{The first author was supported by by the National Research, Development and Innovation Fund
  of Hungary, financed under the Thematic Excellence Programme funding scheme, Project no.\
  ED\_18-1-2019-0030 (Application-specific highly reliable IT solutions). The second author was
  supported by the European Union, co-financed by the European Social Fund
  (EFOP-3.6.3-VEKOP-16-2017-00002).} 




\begin{abstract}
\noindent
Higher inductive-inductive types (HIITs) generalize inductive types of
dependent type theories in two ways. On the one hand they allow the
simultaneous definition of multiple sorts that can be indexed over
each other. On the other hand they support equality constructors,
thus generalizing higher inductive types of homotopy type
theory. Examples that make use of both features are the Cauchy real
numbers and the well-typed syntax of type theory where conversion
rules are given as equality constructors. In this paper we propose a
general definition of HIITs using a small type theory, named the
theory of signatures. A context in this theory encodes a HIIT by
listing the constructors. We also compute notions of induction and
recursion for HIITs by using variants of syntactic logical relation
translations. Building full categorical semantics and constructing
initial algebras is left for future work. The theory of HIIT
signatures was formalised in Agda together with the syntactic
translations. We also provide a Haskell implementation, which takes
signatures as input and outputs translation results as valid Agda
code.
\end{abstract}

\maketitle

\section{Introduction}
\label{sec:intro}

Many dependent type theories support some form of inductive types. An
inductive type is given by its constructors, along with an elimination
principle which expresses that it is enough to consider the
constructors when defining a function out of the inductive type.

For example, the inductive type of natural numbers $\Nat$ is given by
the constructors $\zero:\Nat$ and $\suc:\Nat \ra \Nat$, and has the
well-known induction principle:
\[
  \Elim\Nat:(P:\Nat \ra \mathsf{Type})(pz: P\,\zero)\big(ps:(n:\Nat)\ra P\,n\ra P\,(\suc\,n)\big)(n:\Nat)\ra P\,n
\]
\noindent $P$ is a family of types (i.e.\ a proof-relevant predicate) over natural
numbers, which is called the \emph{induction motive}. The arguments $pz$ and
$ps$ are called the \emph{induction methods}. The behavior of induction is
described by a \emph{computation rule} ($\beta$-rule) for each constructor
and induction method:
\begin{alignat*}{5}
  & \Elim\Nat\,P\,pz\,ps\,\zero && \equiv pz \\
  & \Elim\Nat\,P\,pz\,ps\,(\suc\,n) && \equiv ps\,n\,(\Elim\Nat\,P\,pz\,ps\,n)
\end{alignat*}

Indexed families of types can be also considered, such as
length-indexed vectors of $A$-elements $\mathsf{Vec}_A: \Nat \ra
\mathsf{Type}$. Mutual inductive types are yet another generalization,
but they can be reduced to indexed families where indices classify
constructors for each mutual type.  Inductive-inductive types
\cite{forsberg-phd} are mutual definitions where this reduction does
not work: here a type can be defined together with a family indexed
over it. An example is the following fragment of a well-typed syntax
of a type theory, where the second $\Ty$ type constructor is indexed
over $\Con$, but constructors of $\Con$ also refer to $\Ty$:
\begin{alignat*}{3}
  & \Con && : \mathsf{Type} && \text{contexts} \\
  & \Ty  && : \Con \ra \mathsf{Type} && \text{types in contexts} \\
  & \bullet && : \Con && \text{constructor for the empty context} \\
  & \blank\rhd\blank && : (\Gamma:\Con)\ra\Ty\,\Gamma\ra\Con && \text{constructor for context extension} \\
  & \iota && : (\Gamma:\Con)\ra\Ty\,\Gamma && \text{constructor for a base type} \\
  & \Pi && : (\Gamma:\Con)(A:\Ty\,\Gamma)\ra\Ty\,(\Gamma\rhd A)\ra\Ty\,\Gamma \hspace{1em} && \text{constructor for dependent functions}
\end{alignat*}
There are two eliminators for this type: one for $\Con$ and one for
$\Ty$. Both take the same arguments: two motives ($P:\Con\ra\mathsf{Type}$ and
$Q:(\Gamma:\Con)\ra P\,\Gamma\ra\Ty\,\Gamma\ra\mathsf{Type}$) and four methods
(one for each constructor, which we omit).
\begin{alignat*}{6}
  & \Elim\Con && : (P:\dots)(Q:\dots)\ra\dots\ra(\Gamma:\Con) && \ra P\,\Gamma \\
  & \Elim\Ty  && : (P:\dots)(Q:\dots)\ra\dots\ra(A:\Ty\,\Gamma) && \ra Q\,\Gamma\,(\Elim\Con\,\Gamma)\,A
\end{alignat*}
Note that the type of $\Elim\Ty$ refers to $\Elim\Con$; for this reason
this elimination principle is sometimes called ``recursive-recursive''
(analogously to ``inductive-inductive'').

Higher inductive types (HITs, \cite[Chapter 6]{HoTTbook}) generalize inductive
types in a different way: they allow constructors expressing equalities of
elements of the type being defined. This enables, among others, the definition
of types quotiented by a relation. For example, the type of integers $\Int$ can
be given by a constructor $\mathsf{pair}:\Nat\ra\Nat\ra\Int$ and an equality
constructor $\mathsf{eq}:(a\,b\,c\,d:\Nat)\ra a+d=_\Nat b+c\ra
\mathsf{pair}\,a\,b=_{\Int}\mathsf{pair}\,c\,d$ targetting an equality of
$\Int$. The eliminator for $\Int$ expects a motive $P:\Int\ra\mathsf{Type}$, a
method for the $\mathsf{pair}$ constructor $p:(a\,b:\Nat)\ra
P\,(\mathsf{pair}\,a\,b)$ and a method for the equality constructor
$\mathsf{path}$. This method is a proof that given $e:a+d=_\Nat b+c$, $p\,a\,b$
is equal to $p\,c\,d$ (the types of which are equal by $e$). Thus the method for
the equality constructor ensures that all functions defined from the quotiented
type respect the relation. Since the integers are supposed to be a set (which
means that any two equalities between the same two integers are equal), we would
need an additional higher equality constructor
$\mathsf{trunc}:(x\,y:\Int)\ra(p\,q:x=_\Int y)\ra p=_{x=_\Int y} q$.  HITs may
have constructors of iterated equality types as well. With the view of types as
spaces in mind, point constructors add points to spaces, equality constructors
add paths and higher constructors add homotopies between higher-dimensional paths.

Not all constructor expressions make sense. For example \cite[Example
  6.13.1]{HoTTbook}, given an $f:(X:\mathsf{Type})\ra X\ra X$, suppose that an
inductive type $\mathsf{Ival}$ is generated by the point constructors
$\mathsf{a}:\mathsf{Ival}$, $\mathsf{b}:\mathsf{Ival}$ and a path
constructor $\sigma:f\,\mathsf{Ival}\,\mathsf{a}
=_{\mathsf{Ival}}f\,\mathsf{Ival}\,\mathsf{b}$. The eliminator for this type
should take a motive $P:\mathsf{Ival}\ra\mathsf{Type}$, two methods $p_a :
P\,\mathsf{a}$ and $p_b : P\,\mathsf{b}$, and a path connecting
elements of $P\,(f\,\mathsf{Ival}\,\mathsf{a})$ and
$P\,(f\,\mathsf{Ival}\,\mathsf{b})$. However it is not clear what these
elements should be: we only have elements $p_a:P\,\mathsf{a}$ and
$p_b:P\,\mathsf{b}$, and there is no way in general to transform these
to have types $P\,(f\,\mathsf{Ival}\,\mathsf{a})$ and
$P\,(f\,\mathsf{Ival}\,\mathsf{b})$.

Another invalid example is an inductive type $\mathsf{Neg}$ with a constructor
$\con:(\mathsf{Neg} \ra \bot) \ra \mathsf{Neg}$ where $\bot$ is the empty
type. An eliminator for this type should (at least) yield a projection function
$\proj: \mathsf{Neg} \ra (\mathsf{Neg} \ra \bot)$. Given this, we can define $u
:\equiv \con\,(\lambda x . \proj\, x\,x):\mathsf{Neg}$ and then derive $\bot$ by
$\proj\,u\,u$. The existence of $\mathsf{Neg}$ would make the type theory
inconsistent. A common restriction to avoid such situations is \emph{strict
  positivity}. It means that the type being defined cannot occur on the left
hand side of a function arrow in a parameter of a constructor. This excludes the
above constructor $\con$.

In this paper we propose a notion of signatures for higher inductive-inductive
types (HIITs) which includes the above valid examples and excludes the invalid
ones. Our signatures allow any number of inductive-inductive type constructors,
possibly infinitary higher constructors of any dimension and restricts
constructors to strictly positive ones. It also allows equalities between type
constructors, free usage of $\J$ (path induction) and $\refl$ in HIIT
signatures, and allows mixing type, point and path constructors in any order.

The core idea is to represent HIIT specifications as contexts in a
domain-specific type theory which we call the \emph{theory of signatures}. Type
formers in the theory of signatures are restricted in order to enforce strict
positivity. For example, natural numbers are defined as the three-element
context
\[
  Nat:\U,\,\,\, zero:\underline{Nat},\,\,\, suc : Nat \ra \underline{Nat}
\]
where $Nat$, $zero$ and $suc$ are simply variable names, and underlining
denotes $\El$ (decoding) for the Tarski-style universe $\U$.

We also show how to derive induction and recursion principles for each
signature. We use variants of \emph{syntactic logical relation translations} to
compute notions of \emph{algebras}, \emph{homomorphisms}, \emph{displayed
  algebras} and \emph{displayed algebra sections}, and then define induction and
recursion in terms of these notions.

To our knowledge, this is the first proposal for a definition of
HIITs. However, we do not provide complete (higher) categorical semantics
for HIITs, nor do we show that initial algebras exist for specified HIITs.

The present paper is an expanded version of our conference paper \cite{hiit}.
In this version, we extend signatures with paths between type constructors, and
in Section \ref{sec:morphisms} we also compute notions of homomorphisms from
signatures. We also explain a coherence problem in interpreting syntaxes of type
theories, and how it influenced the current paper, in Section
\ref{sec:coherence} and Section \ref{sec:categorical}.

\subsection{Overview of the Paper}

We start by describing the theory of HIIT signatures in Section
\ref{sec:signatures}. Here, we also describe the syntax for an external type
theory, which serves as the source of constants which are external to a
signature, like natural numbers in the case of length-indexed vectors. In
Section \ref{sec:general}, we give a general definition of induction and
recursion. In Section \ref{sec:coherence} we explain the choice of using
syntactic translations in the rest of the paper. In Sections \ref{sec:algebras}
to \ref{sec:sections}, we describe four syntactic translations from the theory
of signatures to the syntax of the external type theory, respectively computing
algebras, displayed algebras, homomorphisms, and sections of displayed
algebras. In Section \ref{sec:categorical}, we consider extending the previous
translations with additional components of a categorical semantics
(e.g.\ identity and composition for homomorphisms), and explain why the approach
in this paper does not make this feasible. Section \ref{sec:formalization}
describes the Agda formalization and the Haskell implementation. We conclude in
Section \ref{sec:summary}.

\subsection{Related Work}

Schemes for inductive families are given in
\cite{Dybjer97inductivefamilies,paulinmohring}, and for inductive-recursive
types in \cite{dybjer00ir}. A symmetric scheme for both inductive and
coinductive types is given in \cite{henning}. Basold et al. \cite{niels} define
a syntactic scheme for higher inductive types with only 0-constructors and
compute the types of induction principles. In \cite{nielsmsc} a semantics is
given for the same class of HITs but with no recursive equality
constructors. Dybjer and Moeneclaey define a syntactic scheme for finitary HITs
and show their existence in a groupoid model \cite{moeneclaey}.

Internal codes for simple inductive types such as natural numbers, lists or
binary trees can be given by containers which are decoded to W-types
\cite{abbot05containers}. Morris and Altenkirch \cite{morris09indexed} extend
the notion of container to that of indexed container which specifies indexed
inductive types. Codes for inductive-recursive types are given in
\cite{Dybjer99afinite}. Inductive-inductive types were introduced by Forsberg
\cite{forsberg-phd}. Sojakova \cite{sojakova} defines a subset of HITs called
W-suspensions by a coding scheme similar to W-types. She proves that the
induction principle is equivalent to homotopy initiality.

Quotient types \cite{hofmann95extensional} are precursors of higher inductive
types (HITs). The notion of HIT first appeared in \cite{HoTTbook}, however only
through examples and without a general definition.  Lumsdaine and Shulman give a
general specification of models of type theory supporting higher inductive types
\cite{lumsdaineShulman}. They introduce the notion of cell monad with parameters
and characterize the class of models which have initial algebras for a cell
monad with parameters. \cite{cubicalhits} develop semantics for several HITs
(sphere, torus, suspensions, truncations, pushouts) in certain presheaf toposes,
and extend the syntax of cubical type theory \cite{ctt} with these HITs. Kraus
\cite{krausprop} and Van Doorn \cite{doorn} construct propositional truncation
as a sequential colimit. The schemes mentioned so far do not support
inductive-inductive types.

Cartmell's generalized algebraic theories (GATs) \cite{gat} pioneered a
type-theoretic notion of algebraic signature. GATs can be viewed as a variant of
finitary quotient inductive-inductive signatures (QIITs), although GATs also
support equations between sorts (type constructors), which so far have not been
considered in QIIT and HIT literature.

The article of Altenkirch et al.\ \cite{gabe} gives specification and semantics
of QIITs in a set-truncated setting. Signatures are given as lists of functors
which can be interpreted as complete categories of algebras, and completeness is
used to talk about notions of induction and recursion. However, no strict
positivity restriction is given, nor a construction of initial algebras.

Closely related to the current work is the paper by the current
authors and Altenkirch \cite{kaposi2019constructing}, which also
concerns QIITs. There, signatures for QIITs are essentially a
restriction of the signatures given here, but in contrast to the
current work, the restricted set-truncated setting enables building
initial algebras and detailed categorical semantics.

The logical predicate syntactic translation was introduced by Bernardy et
al.\ \cite{bernardy2010parametricity}. The idea that a context can be seen as a
signatures and the logical predicate translation can be used to derive the types
of induction motives and methods was described in \cite[Section
  5.3]{ttintt}. Logical relations are used to derive the computation rules in
\cite[Section 4.3]{kaposi-phd}, but only for closed QIITs. Syntactic
translations in the context of the calculus of inductive constructions are
discussed in \cite{next700}. Logical relations and parametricity can also be
used to justify the existence of inductive types in a type theory with an
impredicative universe \cite{atkey}.

\section{Signatures for HIITs}
\label{sec:signatures}

In this section we define signatures for HIITs. First, we list
the main considerations behind our definition.

\begin{itemize}
\item\emph{Ubiquitous type dependencies.}
Recall the inductive-inductive $\Con$-$\Ty$ example from Section
\ref{sec:intro}: there, types of constructors may refer back to previous
constructors. Additionally, $\Ty$ is indexed over the previously declared $\Con$
type constructor. This suggests that we should not attempt to stratify
signatures, and instead use a fully dependent type theory. At this level of
generality, stratification seems to complicate matters and remove the syntax
further from familiar type theories.

\item\emph{Referring to external types.}
We would like to mention types which are external to the signature. For example,
length-indexed vectors refer to natural numbers which are supposed to already
exist. Hence, we also assume a syntax for an \emph{external type theory}, which
is the source of such types, and constructions in the theory of signatures may
depend on a context in the external type theory.

\item\emph{Strict positivity.}
In prior literature, schemes for inductive types usually include
structural restrictions for this. In our case, a \emph{universe} is
used to make size restrictions which also entail strict positivity.

\item\emph{Iterated equalities and type constructor equalities, with path
induction.} We support iterated equalities by closing the universe under
equality type formation of point and path constructors, together with a
standard (although size-restricted) definition of path induction. We also
introduce an additional type former for equalities between type constructors.

\end{itemize}


\subsection{Theory of Signatures}
\label{sec:tos}

\begin{figure}

(1) Contexts and variables

\[
\begin{gathered}
  \infer{\hat{\Gamma}\vdash\cdot}{\vdashh\,\hat{\Gamma}}
\end{gathered}
\hspace{2em}
\begin{gathered}
  \infer{\hat{\Gamma}\vdash\Delta,x:A}{\targetass\Delta\vdash A}
\end{gathered}
\hspace{2em}
\begin{gathered}
  \infer{\targetass\Delta,x:A\vdash x : A}{\targetass\Delta\vdash A}
\end{gathered}
\hspace{2em}
\begin{gathered}
  \infer{\targetass\Delta,y:B\vdash x : A}{\targetass\Delta\vdash x : A && \targetass\Delta\vdash B}
\end{gathered}
\]

\vspace{0.5em}
(2) Universe

\[
\begin{gathered}
  \infer{\targetass\Delta \vdash \U}{\targetass\vdash\Delta}
\end{gathered}
\hspace{2em}
\begin{gathered}
  \infer{\targetass\Delta \vdash \underline{a}}{\targetass\Delta \vdash a : \U}
\end{gathered}
\]

\vspace{0.5em}
(3) Inductive parameters

\[
\begin{gathered}
  \infer{\targetass\Delta \vdash (x:a)\ra B}{\targetass\Delta \vdash a : \U && \targetass\Delta,x:\underline{a} \vdash B}
\end{gathered}
\hspace{2em}
\begin{gathered}
  \infer{\targetass\Delta \vdash t\, u : B[x \mapsto u]}{\targetass\Delta \vdash t : (x:a)\ra B && \targetass\Delta \vdash u : \underline{a}}
\end{gathered}
\]

\vspace{0.5em}
(4) Paths between point and path constructors

\[
\begin{gathered}
  \infer{\targetass\Delta \vdash t=_a u : \U}{\targetass\Delta \vdash a : \U && \targetass\Delta \vdash t : \underline{a} && \targetass\Delta \vdash u : \underline{a}}
\end{gathered}
\hspace{2em}
\begin{gathered}
  \infer{\targetass\Delta \vdash \refl : \underline{t=_a t}}{\targetass\Delta \vdash t : \underline{a}}
\end{gathered}
\]

\[
\infer{\targetass\Delta \vdash \J_{a\,t\,\,(x.z.p)}\,pr\,_u\,eq : \underline{p[x\mapsto u, z\mapsto eq]}}
      {\begin{array}{l l l}
          & \targetass\Delta,x:\underline{a},z:\underline{t=_a x}\vdash p : \U & \targetass\Delta \vdash u : \underline{a} \\
          \targetass\Delta \vdash t : \underline{a} \hspace{1.5em} & \targetass\Delta \vdash pr : \underline{p[x\mapsto t, z\mapsto \refl]} \hspace{1.5em} & \targetass\Delta \vdash eq : \underline{t=_a u}
      \end{array}}
      \]

\[
\infer{\targetass\Delta \vdash \J\beta_{a\,t\,\,(x.z.p)}\,pr:\underline{(\J_{a\,t\,\,(x.z.p)}\,pr\,_t\,\refl) =_{p[x\mapsto t, z\mapsto \refl]} pr}}
      {\targetass\Delta \vdash t : \underline{a}
        && \targetass\Delta,x:\underline{a},z:\underline{t =_a x}\vdash p : \U
        && \targetass\Delta \vdash pr : \underline{p[x\mapsto t, z\mapsto \refl]}
      }
\]

\vspace{0.5em}
(5) Paths between type constructors

\[
\begin{gathered}
  \infer{\targetass\Delta \vdash a=_\U b}{\targetass\Delta \vdash a : \U && \targetass\Delta \vdash b : \U}
\end{gathered}
\hspace{2em}
\begin{gathered}
  \infer{\targetass\Delta \vdash \refl : a=_\U a}{\targetass\Delta \vdash a : \U}
\end{gathered}
\]

\[
\infer{\targetass\Delta \vdash \J_{a\,\,(x.z.p)}\,pr\,_b\,eq : \underline{p[x\mapsto b, z\mapsto eq]}}
      {\begin{array}{l l l}
          & \targetass\Delta,x:\U,z:a=_\U x\vdash p : \U & \targetass\Delta \vdash b : \U \\
          \targetass\Delta \vdash a : \U \hspace{1.5em} & \targetass\Delta \vdash pr : \underline{p[x\mapsto a, z\mapsto \refl]} \hspace{1.5em} & \targetass\Delta \vdash eq : a=_\U b
      \end{array}}
      \]

\[
\infer{\targetass\Delta \vdash \J\beta_{a\,\,(x.z.p)}\,pr:\underline{(\J_{a\,\,(x.z.p)}\,pr\,_a\,\refl) =_{p[x\mapsto a, z\mapsto \refl]} pr}}
      {\targetass\Delta \vdash a : \U
        && \targetass\Delta,x:\U,z:t =_\U x\vdash p : \U
        && \targetass\Delta \vdash pr : \underline{p[x\mapsto a, z\mapsto \refl]}
      }
\]
\caption{The theory of HIIT signatures, parts (1)--(5). Weakenings are
  implicit, we assume fresh names everywhere and consider
  $\alpha$-convertible terms equal. The $\hat{\Gamma;}$ assumptions
  are only used in parts (6)--(7), see Figure \ref{sigrules2}.}
\label{sigrules1}
\end{figure}

\begin{figure}

(6) External parameters

\[
\begin{gathered}
  \infer{\hat{\Gamma}\semicol\Delta \vdash (\hat{x}\in \hat{A})\ra B}{\hat{\Gamma}\vdashh\hat{A}\in\Type_{\hat{0}} && \hat{\Gamma}\semicol\vdash\Delta && \hat{(\hat{\Gamma}\commah\hat{x}\in \hat{A})}\semicol \Delta \vdash B}
\end{gathered}
\hspace{2em}
\begin{gathered}
  \infer{\hat{\Gamma}\semicol\Delta \vdash t\, \hat{u} : B[\hat{x}\mapsto \hat{u}]}{\hat{\Gamma}\semicol\Delta \vdash t : (\hat{x}\in \hat{A})\ra B && \hat{\Gamma}\vdashh \hat{u} \in \hat{A}}
\end{gathered}
\]

\vspace{0.5em}
(7) Infinitary parameters

\[
\begin{gathered}
  \infer{\hat{\Gamma}\semicol\Delta \vdash (\hat{x}\in \hat{A})\ra b : \U}{\hat{\Gamma}\vdashh\hat{A}\in\Type_{\hat{0}} && \hat{\Gamma}\semicol\vdash\Delta && (\hat{\Gamma}\commah\hat{x}\in \hat{A})\semicol \Delta \vdash b : \U}
\end{gathered}
\hspace{2em}
\begin{gathered}
  \infer{\hat{\Gamma}\semicol\Delta \vdash t\, \hat{u} : \underline{b[\hat{x}\mapsto \hat{u}]}}
        {\hat{\Gamma}\semicol\Delta \vdash t : \underline{(\hat{x}\in \hat{A})\ra b} && \hat{\Gamma}\,\hat{\vdash}\, \hat{u} \in \hat{A}}
\end{gathered}
\]
\caption{The theory of HIIT signatures, parts (6)--(7). Parts (1)--(5)
  are given in Figure \ref{sigrules1}.}
\label{sigrules2}
\end{figure}

We list typing rules for the theory of signatures in Figures \ref{sigrules1} and \ref{sigrules2}. We consider the
following judgments:
\begin{alignat*}{4}
  & \hat{\Gamma}\vdash\Delta && \text{$\Delta$ is a context in the external context $\hat{\Gamma}$}\\
  & \hat{\Gamma}\semicol\Delta\vdash A && \text{$A$ is a type in context $\Delta$ and external context $\hat{\Gamma}$} \\
  & \hat{\Gamma}\semicol\Delta\vdash t : A \hspace{3em} && \text{$t$ is a term of type $A$ in context $\Delta$ and external context $\hat{\Gamma}$}
\end{alignat*}

We have the convention that constructions in the external type theory are
notated in {\color{BrickRed}brick red} color. Although every judgement is valid
up to a context in the external type theory, note that none of the rules in
(1)--(4) depend on or change these assumptions, and even after that, we do not
refer to any particular type former from the external theory. We describe the
external theory in more detail in Section \ref{sec:external}. Also, the rules
presented here are informal and optimized for readability; we describe the Agda
formalizations in Section \ref{sec:formalization}. We explain the rules for the
theory of signatures in order below.

(1) The rules for context formation and variables are standard. We build
signatures in a well-formed external context. We assume fresh names everywhere
to avoid name capture, and leave weakenings implicit.

(2) There is a universe $\U$, with decoding written as an underline
instead of the usual $\El$, to improve readability. With this part of
the syntax, we can already define contexts specifying the empty type,
unit type and booleans, or in general, finite sets of finite sets:
\[
\boldsymbol{\cdot},\,\,\,Empty:\U \hspace{3em} \boldsymbol{\cdot},\,\,\,Unit:\U,\,\,\,tt:\underline{Unit} \hspace{3em} \boldsymbol{\cdot},\,\,\,Bool:\U,\,\,\,true:\underline{Bool},\,\,\,false:\underline{Bool}
\]

(3) We have a function space with small domain and large codomain, which we call
the inductive function space. This can be used to add inductive parameters to all
kinds of constructors. As $\U$ is not closed under this function space, these
function types cannot (recursively) appear in inductive arguments, which ensures
strict positivity. When the codomain does not depend on the domain, $a\ra B$ can
be written instead of $(x:a)\ra B$.

Now we can specify the natural numbers as a context:
\[
\boldsymbol{\cdot},\,\,\,Nat : \U,\,\,\,zero:\underline{Nat},\,\,\,suc:Nat\ra\underline{Nat}
\]
We can also encode inductive-inductive definitions such as the
fragment of the well-typed syntax of a type theory mentioned in the
introduction:
\begin{alignat*}{5}
  & \boldsymbol{\cdot},\,\,\,Con:\U,\,\,\,Ty:Con\ra\U,\,\,\,\bullet:\underline{Con},\,\,\,\blank\rhd\blank:(\Delta:Con)\ra Ty\,\Delta\ra\underline{Con}, \\
  & U : (\Delta:Con)\ra \underline{\Ty\,\Delta},\,\,\,\Pi:(\Delta:Con)(A:Ty\,\Delta)(B:Ty\,(\Delta\rhd A))\ra\underline{Ty\,\Delta}
\end{alignat*}
Note that this notion of inductive-inductive types is more general than the one
considered in previous works \cite{forsberg-phd}, as we allow any number of type
constructors, and arbitrary mixing of type and point constructors.

(4) $\U$ is closed under the equality type, with eliminator $\J$ and a weak
(propositional) $\beta$-rule. Weakness is required because the translations in
Sections \ref{sec:morphisms} and \ref{sec:sections} do not preserve this
$\beta$-rule strictly. We explain this in more detail in Sections
\ref{sec:coherence} and \ref{sec:categorical}. Adding equality to the theory of
signatures allows higher constructors and inductive equality parameters as well. We
can now define the higher inductive circle as the following context:
\[
\boldsymbol{\cdot},\,\,\,S^1:\U,\,\,\,base:\underline{S^1},\,\,\,loop:\underline{base =_{S^1} base}
\]

The $\J$ rule allows constructors to mention operations on paths as well. For
instance, the definition of the torus depends on path composition, which can be
defined using $\J$: given $p:\underline{t=_a u}$ and $q:\underline{u=_a v}$, $p
\sqcdot q$ abbreviates $\J_{a\,u\,x.z.(t=x)}\,p\,_v\,q : \underline{t=_a
  v}$. The torus is given as follows.
\begin{alignat*}{5}
  & \boldsymbol{\cdot},\,\,\,T^2:\U,\,\,\,b : \underline{T^2},\,\,\, p:\underline{b =_{T^2} b},\,\,\,q:\underline{b=_{T^2} b},\,\,\, t:\underline{p\sqcdot q=_{(b=_{T^2} b)} q\sqcdot p}
\end{alignat*}
With the equality type at hand, we can define a full well-typed syntax of type
theory as given e.g.\ in \cite{ttintt} as an inductive type.  Also, see the
examples in the Agda formalization described in Section
\ref{sec:formalization}.

The question may arise whether our $\J$ is sufficient to define all
constructions on paths which we want to express, as we do not have
$\Sigma$-types in signatures, and we only have a weak version of $\Pi$. First,
as far as we know, all path constructions which occur in HITs in the literature
are expressible using our $\J$, so it is certainly adequate in this
sense. Second, we provide a sketch in our Agda formalization (see Section \ref{sec:formalization}) that a Frobenius
variant of $\J$ (first considered by Garner in \cite[p.~13]{garner2009two}) is
derivable from the $\J$ given in Figure \ref{sigrules1}, which suggests that our
$\J$ does not lose expressiveness because of the lack of $\Sigma$ and
$\Pi$-types.

(5) There is a different equality type former, which can be used to express
paths between type constructors. To our knowledge, this has not been considered
in previous HIT literature, although Cartmell considered equations between sorts
in generalized algebraic theories \cite{gat}. For an example, paths between type
constructors allow a compact definition of integers:
\begin{alignat*}{5}
  & \boldsymbol{\cdot},\,\,Int : U,\,\,zero : \underline{Int},\,p : Int =_\U Int
\end{alignat*}
Here, successor and predecessor functions can be recovered from the $p : Int
=_\U Int$ equality, by transporting an $Int$ along $p$ or $p^{-1}$. This
signature could be unfolded to a larger one, by replacing $p : Int =_\U Int$
with an explicit $suc : Int \ra \underline{Int}$ constructor and additional
constructors expressing that $suc$ is an equivalence. The definition of $Int$
with a successor equivalence was previously suggested by Altenkirch and Pinyo
\cite{pinyo2018integers} and Cavallo and M\"ortberg \cite{cavallointegers}.

For another example, we may have a HIIT definition for a type theory where
Russell-style universes are compactly specified with an equality:
\begin{alignat*}{5}
  & ...,\,\, russell : (\Gamma : Con)\ra Tm\,\Gamma\,U =_\U Ty\,\Gamma,\,\, ...
\end{alignat*}

In the presence of the univalence axiom, type constructor equations
can be always equivalently represented using (4) path constructors, by
adding explicit equivalences to a signature. In this case, type
constructor equations serve as a shorthand for equivalences.

So far we were only able to define closed HIITs, which do not refer to external
types. We add rules which include external types into signatures. A context
$\Delta$ for which $\hat{\Gamma}\vdash\Delta$ holds can be seen as a
specification of an inductive type which depends on an external $\hat{\Gamma}$
signature. For example, in the case of lists for arbitrary external element
types, $\hat{\Gamma}$ will be $\hat{A}\in\Type_{\hat{0}}$.

(6) is a function space where the domain is a type in the external theory. We
distinguish it from (3) by using red brick color in the domain specification. We
specify lists and integers as follows, with integers now given as quotients of
pairs of natural numbers:
\begin{alignat*}{4}
  & \hat{A}\in\Type_{\hat{0}} && \vdash\,\boldsymbol{\cdot},\,\, && List:\U,\,\,\,nil:\underline{List},\,\,\,cons:(\hat{x}\in \hat{A})\ra List\ra\underline{List} \\
  & \hat{\Gamma} && \vdash\,\boldsymbol{\cdot},\,\,&& Int:\U,\,\,\,pair:(\hat{x}\,\hat{y}\in\hat{Nat})\ra\underline{Int},\,\,\, \\
  & && && eq:(\hat{a}\,\hat{b}\,\hat{c}\,\hat{d}\in \hat{Nat})(\hat{p}\in \hat{a}\hat{\mathbin{+}}\hat{d}\eqh_{\hat{Nat}} \hat{b}\hat{\mathbin{+}}\hat{c})\ra \underline{pair\,\hat{a}\,\hat{b}=_{Int}\mathsf{pair}\,\hat{c}\,\hat{d}}, \\
  & && && trunc:(x y : Int)(p\,q : a=_{Int} b)\ra \underline{p=_{x=_{Int} y} q}
\end{alignat*}
In the case of integers, $\hat{\Gamma}$ is
$\hat{Nat}\in\Type_{\hat{0}}\hat{,}\,\blank\hat{\mathbin{+}}\blank\in\hat{Nat}\rah
\hat{Nat}\rah\hat{Nat}$, or alternatively, we could require natural
numbers in the external theory. As another example, propositional
truncation for a type $\hat{A}$ is specified as follows.
\[
\hat{A}\in\Type_{\hat{0}}\vdash\, \boldsymbol{\cdot},\,\,\,tr:\U,\,\,\,emb : (\hat{x}\in \hat{A})\ra \underline{tr},\,\,\,eq:(x\,y:tr)\ra \underline{x=_{tr} y}
\]
The smallness of $\hat{A}$ is required in (6). It is possible to generalize
signatures to arbitrary universe levels, but it is not essential to the current
development.

Note that we can assume arbitrary structures in the external $\hat{\Gamma}$
context, which in particular allows us to specify HIITs depending on other
(external) HIITs. We can do this by first specifying a HIIT, then using the
translations in Sections \ref{sec:algebras}-\ref{sec:sections} to compute
notions of algebras and induction, then assume the HIIT in the external context
of another HIIT signature.

The (6) function space preserves strict positivity, since in the external theory
there is no way to recursively refer to the inductive type \emph{being
  defined}. The situation is analogous to the case of $W$-types
\cite{abbot05containers}, where shapes and positions can contain arbitrary types
but they cannot recursively refer to the $W$-type being defined. This setup rules
out some signatures; for example, rose trees cannot be specified as follows, because
we cannot apply the external $\hat{List}$ to the inductive $T$:
\[
\hat{A : \Type_0,\,\,List : \Type_0 \ra \Type_0} \vdash
     \boldsymbol{\cdot},\,\,T : \U,\,\,node : \hat{A} \ra \hat{List}\,T \ra \underline{T}
\]
Analogously nested HIT examples are the ``hubs and spokes'' definitions in
\cite[Section 6.7]{HoTTbook}. To allow such definitions, we would have to
analyze external constructions to check whether they preserve strict
positivity. This is out of the scope of the current paper.

(7) $\U$ is also closed under a function space where the domain is an
external type and the codomain is a small source theory type. We
overload the application notation for external parameters, as it is
usually clear from context which application is meant. The rules allow
types with infinitary constructors, for example, trees branching by a
possibly infinite external type $\hat{A}$:
\[
\hat{A}\in\Type_{\hat{0}}\vdash\,\boldsymbol{\cdot},\,\,\,T:\U,\,\,\,leaf:\underline{T},\,\,\,node:((\hat{x}\in \hat{A})\ra T)\ra\underline{T}
\]
Here, $node$ has a function type (3) with a function type (7) in the
domain. More generally, we can define $W$-types
\cite{abbot05containers} as follows. $\hat{S}$ describes the
``shapes'' of the constructors and $\hat{P}$ the ``positions'' where
recursive arguments can appear.
\[
\hat{S : \Type_0, P : S \ra \Type_0} \vdash\,\boldsymbol{\cdot},\,\,\,W:\U,\,\,\,sup: (\hat{s} \in \hat{S})\ra((\hat{p} \in \hat{P}\,\hat{s})\ra W)\ra \underline{W}
\]
For a more complex infinitary example, see the definition of Cauchy
reals in \cite[Definition 11.3.2]{HoTTbook}. It can be also found as
an example file in our Haskell implementation.

Note that we do not include a $\lambda$ for infinitary function types,
although we could possibly use it when describing paths between such
functions. The reason is that $\lambda$ can be always represented in
signatures by introducing additional function parameters which are
constrained by pointwise equalities. For example, consider adding the
following path constructor to the previously described
$\hat{A}$-branching trees:
\[
eq : \underline{node\,(\lambda\,\hat{x}.\,leaf) =_T leaf}
\]
This can be rewritten without $\lambda$ as follows:
\[
eq : (f : \hat{A}\ra T)\ra ((\hat{x : A})\ra f \hat{x} =_T leaf)\ra\underline{node\,f =_T leaf}
\]
The benefit of supporting $\lambda$ would be less encoding noise in signatures. However,
we observe that such usage of $\lambda$ is rather rare in HITs in practice, and hence omit
$\lambda$ for the sake of simplicity.

The invalid examples $\mathsf{Ival}$ and $\mathsf{Neg}$ from Section
\ref{sec:intro} cannot be encoded by the theory of signatures. For $\mathsf{Ival}$,
we can go as far as
\[
\boldsymbol{\cdot},\,\,\,{Ival}:\U,\,\,\,a:\underline{{Ival}},\,\,\,b:\underline{{Ival}},\,\,\,\sigma:\underline{?  =_{{Ival}} ?}.
\]
The first argument of the function
$\hat{\hat{f}\in(\hat{X}:\Type)\rah\hat{X}\rah\hat{X}}$ is an external type, but we
only have ${Ival}:\U$ in the theory of signatures. $\mathsf{Neg}$ cannot be typed
because the first parameter of the constructor $\mathsf{con}$ is a function from
a small type to an external type, and no such functions can be formed.

\subsection{External Type Theory}
\label{sec:external}

The external syntax serves two purposes: it is a source of types external to a
HIIT signature, and it also serves as the target for the syntactic translations
described in Sections \ref{sec:algebras} to \ref{sec:sections}. It is not
essential that we use the same theory for both purposes; we do so only to
simplify the presentation by skipping an additional translation or embedding
step. Also, we do not specify the external theory in formal detail, since it is
a standard type theory. We only make some assumptions about supported type
formers. We generally keep the notation close to Agda, and use
{\color{BrickRed}brick red} color to distinguish from constructions in the
theory of signatures.

There is a cumulative Russell-style hierarchy of universes $\Type_{\hat{i}}$,
with universes closed under $\Pi$, $\Sigma$, equality and unit
types. Importantly, we do not assume uniqueness of identity proofs.

The unit type is denoted $\hat{\top}$ with constructor $\tt$.

Dependent function space is denoted $\hat{(\hat{x}\in \hat{A})\rah \hat{B}}$. We
write $\hat{A}\rah \hat{B}$ if $\hat{B}$ does not depend on $\hat{x}$, and $\rah
$ associates to the right, $\hat{(\hat{x}\in \hat{A})(\hat{y}\in \hat{B})\rah
  \hat{C}}$ abbreviates $\hat{(\hat{x}\in \hat{A})\rah (\hat{y}\in \hat{B})\rah
  \hat{C}}$ and $\hat{(\hat{x}\,\hat{y}\in \hat{A})\rah \hat{B}}$ abbreviates
$\hat{(\hat{x}\in \hat{A})(\hat{y}\in \hat{A})\rah \hat{B}}$. We write
$\hat{\lambda x. t}$ for abstraction and $\hat{t}\,\hat{u}$ for left-associative
application.

$\hat{(\hat{x}\in \hat{A})\timesh \hat{B}}$ stands for $\Sigma$ types, $\hat{A}\timesh
\hat{B}$ for the non-dependent version. We sometimes use a short re-associated
notation for left-nested iterated $\Sigma$ types, for example $\hat{(A :
  \Type_{\hat{0}})\times A \times A}$ may stand for the left-nested $\hat{(x :
  (A : \Type_{\hat{0}})\times A) \times \proj_1\,A}$. The constructor for
$\Sigma$ is denoted $\hat{(t,\,u)}$ with eliminators $\hat{\proj_1}$
and $\hat{\proj_2}$. Both $\Pi$ and $\Sigma$ have definitional $\beta$ and
$\eta$ rules.

The equality type for a type $\hat{A}$ and elements $\hat{t}\in \hat{A}$,
$\hat{u}\in \hat{A}$ is denoted $\hat{t}\eqh_{\hat{A}}\hat{u}$, and we have the
constructor $\hat{\refl}_{\hat{t}}$ and the eliminator $\hat{\J}$ with
definitional $\beta$-rule. The notation is
$\hat{\J}_{\hat{A}\,\hat{t}\,\hat{P}}\,\hat{pr}\,_{\hat{u}}\,\hat{eq}$ for
$\hat{t}\in \hat{A}$, $\hat{P}\in (\hat{x}\in \hat{A})\rah
\hat{t}\eqh_\hat{A}\hat{x}\rah \Type_{\hat{i}}$, $\hat{pr}\in
\hat{P}\,\hat{t}\,\hat{\refl}$ and $\hat{eq} \in
\hat{t}\eqh_\hat{A}\hat{u}$. Sometimes we omit parameters in subscripts.

We will use the following functions defined using $\hat{\J}$ in the
standard way. We write $\TR_{\hat{P}}\,\hat{e}\,\hat{t}\in
\hat{P}\,\hat{v}$ for transport of $\hat{t} \in \hat{P}\,\hat{u}$
along $\hat{e} \in \hat{u}\eqh \hat{v}$ and $\hat{\coe\,e\,t : B}$ for
coercion of $\hat{t:A}$ along $\hat{e : A =_\Type B}$. We write
$\ap\,\hat{f}\,\hat{e}\in\hat{f}\,\hat{u}\eqh\hat{f}\,\hat{v}$ where
$\hat{f}:\hat{A}\rah\hat{B}$ and $\hat{e}\in\hat{u}\eqh\hat{v}$, also
$\hat{\apd\,\hat{f}\,\hat{e} \in
  \TR_{\hat{P}}\,\hat{e}\,(\hat{f}\,\hat{u}) \eqh \hat{f}\,\hat{v}}$
where $\hat{\hat{f}\in(\hat{x}\in \hat{A})\rah \hat{B}}$ and $\hat{e}
\in \hat{u}\eqh \hat{v}$. Borrowing notation from the homotopy type
theory book \cite{HoTTbook}, we write $\hat{p \sqcdot q}$ for
transitivity and $\hat{p^{-1}}$ for symmetry. We also make use of the
groupoid law $\hat{\inv\,p : p^{-1}\sqcdot p = \refl}$.


\section{General Definitions for Induction and Recursion}
\label{sec:general}

Armed with a definition for HIIT signatures, we would like to have of notions of
\emph{induction} and \emph{recursion} for each signature. However, instead of
trying to directly extract them from signatures, it is more helpful to have more
fundamental (categorical) semantic concepts: \emph{algebras},
\emph{homomorphisms}, \emph{displayed algebras} and \emph{sections of displayed
  algebras}. Then, we can express induction and recursion using these.

Let us first consider natural numbers, and see how the usual
definition of induction arises. For $\Nat$, algebras are simply a
triple consisting of a type, a value and an endofunction. Below, we
leave universe indices explicit, and we use the notation of the
external type theory described in Section \ref{sec:external}.
\begin{alignat*}{5}
& \Alg : \mathsf{Type} \\
& \Alg \equiv (N : \mathsf{Type}) \times N \times (N \ra N)
\end{alignat*}

\noindent Displayed $\Nat$-algebras (sometimes called fibered algebras, as in
\cite{sojakova}) are likewise triples, but each component depends on the
corresponding components of a $\Nat$-algebra. We borrow the term ``displayed''
from Ahrens and Lumsdaine \cite{displayedCategories}, as our displayed algebras
generalize their displayed categories.
\begin{alignat*}{5}
& \DisplayedAlg : \Alg \ra \mathsf{Type} && \\
& \DisplayedAlg\,(N,\,z,\,s) \equiv && \\
&  \hspace{3em} (N^D : N \ra \mathsf{Type})\times (z^D : N^D\,z)\times ((n : N)\ra N^D\,n\ra N^D\,(s\,n))
\end{alignat*}

\noindent  Homomorphisms, as usual in mathematics, are structure-preserving functions:
\begin{alignat*}{5}
& \Morphism : \Alg \ra \Alg \ra \mathsf{Type} && \\
& \Morphism\,(N_0,\,z_0,\,s_0)\,(N_1,\,z_1,\,s_1) \equiv && \\
&  \hspace{3em} (N^M : N_0 \ra N_1)\times (z^M : N^M\,z_0 = z_1)\times ((n : N_0)\ra N^M\,(s_0\,n) = s_1\,(N^M\,n))
\end{alignat*}

\noindent Sections of displayed algebras can be viewed as a dependently typed analogue of homomorphisms:
\begin{alignat*}{5}
& \Section : (\alpha : \Alg) \ra \DisplayedAlg\,\,\alpha \ra \mathsf{Type} && \\
& \Section\,(N,\,z,\,s)\,(N^D,\,z^D,\,s^D) \equiv && \\
&  \hspace{3em} (N^S : (n : N) \ra N^D\,n)\times (z^S : N^S\,z = z^D)\times ((n : N)\ra N^S\,(s\,n) = s^D\,n\,(N^S\,n))
\end{alignat*}

\noindent Now, we can reformulate induction for $\Nat$. First, we
assume that there exists a distinguished $\Nat$-algebra, named
$\Nat^*$. The induction principle for this algebra has the following
type:
\[
 \mathsf{Induction} : (M : \mathsf{\DisplayedAlg}\,\,\Nat^*) \ra \Section\, M
\]
\noindent Unfolding the definitions, it is apparent that this is the same notion
of $\Nat$-induction as we gave before. The initial algebra consists of type and value
constructors, the induction motives and methods are bundled into a displayed
algebra, and as result we get a section, containing an eliminator function together
with its $\beta$-rules. Additionally, we can define recursion using homomorphisms:
\[
\mathsf{Recursion} : (\alpha : \Alg) \ra \Morphism\,\Nat^* \,\alpha
\]
This corresponds to \emph{weak initiality} in the sense of category theory: for
each algebra, there is a morphism from the weakly initial algebra to it. Strong initiality
in the setting of higher inductive types is called \emph{homotopy initiality} \cite{sojakova},
and it is defined as follows for $\Nat$:
\begin{alignat*}{5}
\mathsf{Initiality} : (\alpha : \Alg) \ra \mathsf{isContr}\,(\Morphism\,\Nat^*\,\alpha)
\end{alignat*}

\noindent where $\mathsf{isContr}\,A \equiv (a : A)\times((a' : A)\ra a = a')$. Hence, there is a unique morphism from the initial algebra, but in the setting of homotopy type theory, unique inhabitation can be viewed instead as contractibility.

Observe that the definitions for $\mathsf{Induction}$, $\mathsf{Recursion}$ and
$\mathsf{Initiality}$ need not refer to natural numbers, and can be used
similarly in cases of other structures. Thus, the task in the following is to
derive algebras, homomorphisms, displayed algebras and sections from HIIT
signatures, in a way which generalizes beyond the current $\Nat$ example to
indexed types, induction-induction and higher constructors. But even in the
general case, displayed algebras yield induction motives and methods, and
homomorphisms and sections yield a function for each type constructor and a
$\beta$-rule for each point or path constructor.

We compute algebras and the other notions by induction on the syntax
of the theory of signatures. However, first we need to clarify the
formal foundations of these computations.

\section{The Coherence Problem and Syntactic Translations}
\label{sec:coherence}

The next task would be to define a computation which takes as input a
$\hat{\Gamma}\vdash\Delta$ signature, and returns the corresponding
type of algebras in some type theory. This would behave as a
``standard'' model of signatures, which simply maps each construction
in the syntax to its counterpart: types to types, universe to
universe, functions to functions, and so on.

However, it is important to interpret signatures into a theory without
uniqueness of identity proofs (UIP), because we are considering \emph{higher}
inductive types, and hence must remain compatible with higher-dimensional
interpretations. In particular, we need to interpret type constructors in
signatures into type universes which are not truncated to any homotopy level. In
this setting, even the mundane $\alpha : (N : \mathsf{Type}) \times N \times (N
\ra N)$ natural number algebras may have arbitrary higher-dimensional structure.

On first look, we might think that the simplest way to formalize the algebra
interpretation is the following:
\begin{enumerate}
\item Assume as metatheory a type theory without UIP.
\item In this setting, define a formal syntax of the theory of signatures.
\item Give a standard interpretation of signatures into the UIP-free metatheory.
\end{enumerate}

\noindent
It may come as a surprise that realizing the above steps is an
\emph{open problem}, for any syntax of a dependent type theory. We
call the problem of interpreting syntaxes of type theories into a
UIP-free metatheory a \emph{coherence problem}.

This issue appears to arise with all known ways of defining syntaxes
for dependent type theories. Shulman previously discussed this problem
in \cite{hottshouldeat}. It is also related to the problem of
constructing semisimplicial types in homotopy type theory; as
explained in \cite{hottshouldeat} solving our coherence problem
enables a construction of semisimplicial types.

In the following, we first consider the coherence problem in two settings: with
intrinsically typed higher inductive-inductive syntaxes, then with conventional
syntaxes involving preterms and inductively defined typing and conversion
relations. Then, we present syntactic translations as a partial solution to the
coherence problem.

\subsection{Interpreting Intrinsic Syntax}
Following Altenkirch and Kaposi \cite{ttintt}, one might define the
syntax of a type theory as a category with families (CwF)
\cite{dybjer1995internal} extended with additional type formers, which
supports an induction principle. The CwF part provides a calculus and
equational theory for explicit substitutions, upon which one can build
additional type structure. We present an excerpt below:

\begin{alignat*}{5}
  & \Con && : \Set && \text{contexts} \\
  & \Ty  && : \Con\ra\Set && \text{types} \\
  & \Sub  && : \Con\ra\Con\ra\Set && \text{substitutions} \\
  & \Tm  && : (\Gamma:\Con)\ra\Ty\,\Gamma\ra\Set \hspace{2em} && \text{terms} \\
  & \cdot && : \Con && \text{empty context} \\
  & \blank\rhd\blank && : (\Gamma:\Con)\ra\Ty\,\Gamma\ra\Con && \text{context extension} \\
  & \blank[\blank] && : \Ty\,\Delta\ra\Sub\,\Gamma\,\Delta\ra\Ty\,\Gamma && \text{type substitution} \\
  & \id && : \Sub\,\Gamma\,\Gamma && \text{identity substitution} \\
  & [\id] && : A[\id] = A && \text{action of}\,\,\id\,\,\text{on types}\\
  & \blank\circ\blank && : \Sub\,\Theta\,\Delta\ra\Sub\,\Gamma\,\Theta\ra\Sub\,\Gamma\,\Delta && \text{substitution composition} \\
  & \blank[\blank] && : \Tm\,\Delta\,A\ra(\sigma:\Sub\,\Gamma\,\Delta)\ra\Tm\,\Gamma\,(A[\sigma])\hspace{1em} && \text{term substitution} \\
  & ... && \\
  & \U && : \Ty\,\Gamma && \text{universe} \\
  & {\U[]} && : \U[\sigma] = \U && \text{substituting the universe} \\
  & \El && : \Tm\,\Gamma\,\U \ra \Ty\,\Gamma && \text{decoding} \\
  & ... && \\
  & \Pi && : (a:\Tm\,\Gamma\,\U)\ra\Ty\,(\Gamma\rhd \E\,a)\ra\Ty\,\Gamma && \text{functions} \\
  & ... &&
\end{alignat*}

This notion of syntax is much more compact and often more convenient to use
than extrinsic syntaxes. It is in essence ``merely'' a formalization of CwFs,
which are often used in categorical semantics of type theory. However, its
rigorous metatheory is subject to ongoing research (including the current
paper). In a set-truncated setting, the current authors and Altenkirch have
previously developed semantics and constructed initial algebras
\cite{kaposi2019constructing}, but here we need to work in a non-truncated
theory.

\subsubsection{Set-truncation}
Additionally, we prefer to set-truncate the syntax by adding the following
constructors, for reasons shortly explained:
\begin{alignat*}{5}
  & \mathsf{setTy}  && : (A\,B : \Ty\,\Gamma)(p\,q : A = B)\ra p = q \\
  & \mathsf{setTm}  && : (t\,u : \Tm\,\Gamma\,A)(p\,q : t = u)\ra p = q \\
  & \mathsf{setSub} && : (\sigma\,\delta : \Sub\,\Gamma\,\Delta)(p\,q : \sigma = \delta)\ra p = q
\end{alignat*}

\noindent We can omit the rule for contexts, as it is derivable from the above
ones. Set truncation forces definitional equality of the syntax (as defined by
equality constructors in the HIIT signature) to be proof irrelevant. If we omit
set-truncation, then the defined HIIT becomes very different from what we expect
the syntax to be.

We can easily show that the non-truncated syntax does not form sets. For
example, $\U[]$ and $[\id]$ are two proofs of $\U[\id] = \U$, and they are not
forced to be equal. Assuming univalence, we can give a model where types are
interpreted as closed metatheoretic types, $\U$ is interpreted as metatheoretic
$\Bool$, $[\id]$ is interpreted as $\refl : A = A$ for some metatheoretic $A$
type, and $\U[]$ is interpreted as the $\Bool$ negation equivalence, thereby
formally distinguishing $\U[]$ and $[\id]$. Hence, by Hedberg's theorem
\cite{hedberg}, the syntax does not have decidable equality, and hence it does
not support decidable type checking. This also implies that the non-truncated
intrinsic syntax is not constructible from set quotients of extrinsic terms
(since those always form sets). The non-truncated syntax just does not seem to
be a sensible notion. This situation is similar to how categories in homotopy
type theory need to have set-truncated morphisms \cite{ahrens2015univalent}.

Unfortunately, set-truncation makes it impossible to directly interpret
syntactic types as elements of a UIP-free metatheoretic $\mathsf{Type}_i$
universe. This is because we must provide interpretations for all set-truncation
constructors, which amounts to shoving that the interpretations of $\Ty$,
$\Tm$, and $\Sub$ are all sets. However, we cannot show $\mathsf{Type}_i$ to be a
set without UIP.

A possible solution would be to add \emph{all higher coherences} instead of
set-truncating, which would yield something like an ($\omega$, 1)-CwF, but this
is also an open research problem \cite{altenkirch2018towards, finster2019structure}.

\subsection{Interpreting Extrinsic Syntax}
An extrinsic syntax for type theory is defined the following way:
\begin{enumerate}
\item We inductively define a \emph{presyntax}: sets of preterms, pretypes,
      precontexts, and possibly presubstitutions. These are not assumed to
      be well-formed, and only serve as raw material for expressions.
\item We give mutual inductive definitions for the following relations on presyntax:
      well-formedness, typing and conversion.
\end{enumerate}

\noindent
This is the conventional way of presenting the syntax; see
e.g.\ \cite{winterhalter2019eliminating} for a detailed machine-checked
formalization in this style. The main advantage compared to the intrinsic syntax
is that this only requires conservative inductive definitions in the metatheory,
which are also natively supported in current proof assistants, unlike HIITs. The
main disadvantage is verbosity, lower level of abstraction, and a difficulty of
pinning down a notion of model for the syntax.

What about interpreting extrinsic syntax into a UIP-free universe? It is widely
accepted that extrinsic syntaxes have standard interpretations in set-truncated
metatheories, although carrying this out in formal detail is technically
challenging. Streicher's seminal work \cite{streicher2012semantics} laid out a
template for doing this: first we construct a family of partial functions from
the presyntax to the semantic domain, then we prove afterwards that these
functions are total on well-formed input.

However, the coherence problem arises still: it is required that definitional
equality is proof irrelevant. In a type-theoretic setting, this means that we
need to propositionally truncate the conversion relation. This again prevents us
from interpreting the syntax into a UIP-free universe. We have to interpret
definitional equality in the syntax as propositional equality in the metatheory,
but since the former is propositionally truncated, we can only eliminate it into
propositions, and metatheoretic equality types are not generally propositions in
the absence of UIP.

Could we define a conversion relation which is propositional, but not truncated?
For example, conversion could be defined in terms of a deterministic conversion
checking algorithm. But then a complication is that we do not have a proof that
conversion checking is \emph{total} and \emph{stable under substitution}, while
still in the process of defining the syntax.

Alternatively, we could first define a normalization algorithm for an extrinsic
syntax in a UIP-free metatheory, and then try to interpret \emph{only normal
  forms}. Abel, \"Ohman and Vezzosi demonstrated a UIP-free conversion checking
algorithm in type theory \cite{abel2017decidability}, which suggests that
UIP-free normalization may be possible as well. But since this normal form
interpretation is a major technical challenge, and it has not been carried out
yet, we cannot use it to justify constructions in the current paper.

\subsection{Syntactic Translations}
We can circumvent the coherence problem in the following way:
\begin{enumerate}
\item Define a \emph{source} and a \emph{target} syntax in any suitable metatheory,
      where the target theory does not have UIP. The source and target theories
      do not necessarily need to differ.
\item Interpret the source syntax into the target syntax.
\end{enumerate}

\noindent
For extrinsic syntaxes, it is generally understood that a syntactic translation
has to preserve definitional equalities in the source syntax. For intrinsic
syntaxes, preservation of definitional equality is automatically enforced by the
equality constructors. See Boulier et al.\ \cite{next700} for a showcase of
syntactic translations.

Now, we can take the source syntax to be the theory of signatures from Section
\ref{sec:tos}, and the target syntax to be the external syntax from Section
\ref{sec:external}. Truncation in the source syntax is not an issue here,
because the target syntax is likewise truncated.

However, using syntactic translations is also a significant restriction. As
always, we must map equal inputs to equal outputs, but now the notion of
equality for outputs coincides with definitional equality in the syntax of the
target theory, which is far more restrictive than propositional equality. Recall
the weak $\beta$-rule for $\J$ in the theory of signatures in Section
\ref{sec:tos}: if we instead used a strict equality, then the translations in
Sections \ref{sec:morphisms} and \ref{sec:sections} would not work, because they
map $\J_{a\,t\,\,(x.z.p)}\,pr\,_t\,\refl$ and $pr$ to terms which are equal
propositionally, but not definitionally. This restriction also prevents us from
defining more translations which cover other parts of the categorical
semantics, e.g.\ composition of homomorphisms. We return to this topic in
Section \ref{sec:categorical}.

In the following three sections we present syntactic translations yielding
algebras, homomorphisms, displayed algebras and their sections. The presentation
here, like in Section \ref{sec:tos}, is informal and focuses on readability.
In particular, we omit interpretations for substitutions and preservation
proofs for definitional equalities.

\section{Algebras}
\label{sec:algebras}

We use $\blank^\A$ to the denote the translation which computes algebras. It is
specified as follows, for contexts, types and terms in the theory of signatures.

\[
\infer{\hat{\Gamma}\vdashh\Delta^\A\in\Type_{\hat{1}}}{\hat{\Gamma}\vdash\Delta}
\hspace{2em}
\infer{\hat{\Gamma}\vdashh A^\A \in \Delta^\A\rah \Type_{\hat{1}}}{\hat{\Gamma}\semicol\Delta\vdash A}
\hspace{2em}
\infer{\hat{\Gamma}\vdashh t^\A \in \hat{(}\hat{\gamma}\in\Delta^\A\hat{)}\rah  A^\A\,\hat{\gamma}}{\hat{\Gamma}\semicol\Delta\vdash t : A}
\]

The $\blank^\A$ translation is essentially the standard interpretation of
signatures, where every construction in the source syntax is interpreted with a
corresponding {\color{BrickRed}brick red} construction in the target syntax. The
only notable change is in the interpretation of contexts: a source context is
interpreted as an iterated $\Sigma$-type.

\begin{alignat*}{5}
  & (1)\hspace{1em} && \boldsymbol{\cdot}^\A && :\equiv \hat{\top} \\
  & && (\Delta,x:A)^\A && :\equiv \hat{(}\hat{\gamma}\in\Delta^\A\hat{)}\timesh A^\A\,\hat{\gamma} \\
  & && x^\A\,\hat{\gamma} && :\equiv x^{\text{th}}\text{ component in } \hat{\gamma} \\
  & (2) && \U^\A\,\hat{\gamma} && :\equiv \Type_{\hat{0}} \\
  & && (\underline{a})^\A\,\hat{\gamma} && :\equiv a^\A\,\hat{\gamma} \\
  & (3) && ((x:a)\ra B)^\A\,\hat{\gamma} && :\equiv \hat{(}\hat{x}\in a^\A\,\hat{\gamma}\hat{)}\rah  B^\A\,\hat{(}\hat{\gamma}\commah\hat{x}\hat{)} \\
  & && (t\,u)^\A\,\hat{\gamma} && :\equiv \hat{(}t^\A\,\hat{\gamma}\hat{)}\,\hat{(}u^\A\,\hat{\gamma}\hat{)} \\
  & (4) && (t=_a u)^\A\,\hat{\gamma} && :\equiv t^\A\,\hat{\gamma} \eqh u^\A\,\hat{\gamma} \\
  & && (\refl_t)^\S\,\hat{\gamma} && :\equiv \hat{\refl} \\
  & && (\J_{a\,t\,(x.z.p)}\,pr\,_u\,eq)^\A\,\hat{\gamma} && :\equiv \hat{\J}_{\hat{(}a^\A\,\hat{\gamma}\hat{)}\,\hat{(}t^\A\,\hat{\gamma}\hat{)}\,\hat{(}\hat{\lambda} \hat{x}\,\hat{z}.p^\A\,\hat{(}\hat{\gamma}\commah\hat{x}\commah\hat{z}\hat{)}\hat{)}}\,\hat{(}pr^\A\,\hat{\gamma}\hat{)}\,_{\hat{(}u^\A\,\hat{\gamma}\hat{)}}\,\hat{(}eq^\A\,\hat{\gamma}\hat{)} \\
  & && (\J\beta_{a\,t\,(x.z.p)}\,pr)^\A\,\hat{\gamma} && :\equiv \hat{\refl} \\
  & (5) && (a=_\U b)^\A\,\hat{\gamma} && :\equiv a^\A\,\hat{\gamma} \eqh b^\A\,\hat{\gamma} \\
  & && (\refl_a)^\A\,\hat{\gamma} && :\equiv \hat{\refl} \\
  & && (\J_{a\,(x.z.p)}\,pr\,_b\,eq)^\A\,\hat{\gamma} && :\equiv \hat{\J}_{\Type_{\hat{0}}\,\hat{(}a^\A\,\hat{\gamma}\hat{)}\,\hat{(}\hat{\lambda} \hat{x}\,\hat{z}.p^\A\,\hat{(}\hat{\gamma}\commah\hat{x}\commah\hat{z}\hat{)}\hat{)}}\,\hat{(}pr^\A\,\hat{\gamma}\hat{)}\,_{\hat{(}b^\A\,\hat{\gamma}\hat{)}}\,\hat{(}eq^\A\,\hat{\gamma}\hat{)} \\
  & && (\J\beta_{a\,(x.z.p)}\,pr)^\A\,\hat{\gamma} && :\equiv \hat{\refl} \\
  & (6) && ((\hat{x}\in \hat{A})\ra B)^\A\,\hat{\gamma} && :\equiv \hat{(}\hat{x}\in \hat{A}\hat{)}\rah  B^\A\,\hat{\gamma} \\
  & && (t\,\hat{u})^\A\,\hat{\gamma} && :\equiv \hat{(}t^\A\,\hat{\gamma}\hat{)}\,\hat{u} \\
  & (7) && ((\hat{x}\in \hat{A})\ra b)^\A\,\hat{\gamma} && :\equiv \hat{(}\hat{x}\in \hat{A}\hat{)}\rah  b^\A\,\hat{\gamma} \\
  & && (t\,\hat{u})^\A\,\hat{\gamma} && :\equiv \hat{(}t^\A\,\hat{\gamma}\hat{)}\,\hat{u}
\end{alignat*}

For example, $\blank^\A$ acts as follows on signature for circles:
\begin{alignat*}{5}
  & && (\boldsymbol{\cdot},\,S^1 : \U,\,b:\underline{S^1},\,loop: \underline{b=b})^\A \equiv \hat{\hat{\top}\timesh(\hat{S^1}\in\Type_{\hat{0}})\timesh(\hat{b}\in \hat{S^1})\timesh(\hat{loop\in b=b})}
\end{alignat*}

Note that the resulting $\Sigma$ type is left-nested, and we use the reassociated notation for readability. The result could be written without syntactic sugar the following way:
\begin{alignat*}{5}
  & \hat{\Big(\hat{x''}\in\big(\hat{x'}\in(\hat{x}\in\hat{\top})\timesh\Type_{\hat{0}}\big)\timesh\hat{\proj_2}\,\hat{x'}\Big)\timesh(\hat{\proj_2}\,\hat{x''}\eqh \hat{\proj_2}\,\hat{x''})}
\end{alignat*}
We shall keep to the short notation from now on.


\section{Displayed Algebras}
\label{sec:displayed}

The $\blank^\D$ translation computes displayed algebras for a signature, which
can be viewed as bundles of induction motives and methods. $\blank^\D$ is an
unary logical predicate translation over the $\blank^\A$ translation. It is
related to the logical predicate translation of Bernardy et
al.\ \cite{bernardy12parametricity}, but our implementation differs by
interpreting contexts as $\Sigma$-types instead of extended contexts.

We fix a universe level $\hat{i}$ for the translation. For each context
$\Delta$, $\Delta^\D$ is a predicate over the standard interpretation
$\Delta^\A$. For a type $\Delta\vdash A$, $A^\D$ is a predicate over $A^\A$,
which also depends on $\hat{\gamma}\in\Delta^\A$ and a witness of
$\Delta^\D\,\hat{\gamma}$. All of these may refer to a target theory context
$\hat{\Gamma}$.
\[
\infer{\hat{\Gamma}\vdashh\Delta^\D \in \Delta^\A\rah \Type_{\hat{i+1}}}{\hat{\Gamma}\vdash\Delta}
\hspace{2em}
\infer{
  \hat{\Gamma}\vdashh A^\D \in \hat{(}\hat{\gamma}\in\Delta^\A\hat{)}\rah \Delta^\D\,\hat{\gamma}\rah  A^\A\,\hat{\gamma}\rah \Type_{\hat{i+1}}}
      {\hat{\Gamma}\semicol\Delta\vdash A}
\]
For a term $t$, $t^\D$ witnesses that the predicate corresponding to its type
holds for $t^\A$; this can be viewed as a \emph{fundamental theorem} for the
predicate interpretation.

\[
\infer{\hat{\Gamma}\vdashh t^\D \in \hat{(}\hat{\gamma}\in\Delta^\A\hat{)}\hat{(}\hat{\gamma^D}\in\Delta^\D\,\hat{\gamma}\hat{)}\rah  A^\D\,\hat{\gamma}\,\hat{\gamma^D}\,\hat{(}t^\A\,\hat{\gamma}\hat{)}}{\hat{\Gamma}\semicol\Delta\vdash t : A}
\]
The implementation of $\blank^\D$ is given below. We leave
$\hat{\gamma}$-s mostly implicit, marking only $\hat{\gamma^D}$ witnesses.
\begingroup
\allowdisplaybreaks
\begin{alignat*}{5}
  & (1)\hspace{1em} && \boldsymbol{\cdot}^\D\,\hat{\gamma} && :\equiv \hat{\top} \\
  & && (\Delta,\,x:A)^\D\,(\hat{\gamma}\commah\,\hat{t}) && :\equiv \hat{(}\hat{\gamma^D}\in\Delta^D\,\hat{\gamma}\hat{)}\timesh A^D\,\hat{\gamma^D}\,\hat{t} \\
  & && x^\D\,\hat{\gamma^D} && :\equiv x^{\text{th}}\text{ component in } \hat{\gamma^D} \\
  & (2) && \U^\D\,\hat{\gamma^D}\,\hat{A} && :\equiv \hat{A} \rah  \Type_{\hat{i}} \\
  & && (\underline{a})^\D\,\hat{\gamma^D}\,\hat{t} && :\equiv a^\D\,\hat{\gamma^D}\,\hat{t} \\
  & (3) && ((x:a)\ra B)^\D\,\hat{\gamma^D}\,\hat{f} && :\equiv \hat{(}\hat{x}\in a^\A\,\hat{\gamma}\hat{)}\hat{(}\hat{x^D}\in a^\D\,\hat{\gamma^D}\,\hat{x}\hat{)} \rah  B^\D\,\hat{(}\hat{\gamma}\commah\hat{x}\hat{)}\,\hat{(}\hat{\gamma^D}\commah\hat{x^D}\hat{)}\,\hat{(}\hat{f}\,\hat{x}\hat{)} \\
  & && (t\,u)^\D\,\hat{\gamma^D} && :\equiv \hat{(}t^\D\,\hat{\gamma^D}\hat{)}\,\hat{(} u^\A\,\hat{\gamma}\hat{)}\,\hat{(}u^\D\,\hat{\gamma^D}\hat{)} \\
  & (4) && (t=_a u)^\D\,\hat{\gamma^D}\,\hat{e} && :\equiv \TR_{\hat{(}a^\D\,\hat{\gamma^D}\hat{)}}\,\hat{e}\,\hat{(}t^\D\,\hat{\gamma^D}\hat{)} \eqh u^\D\hat{\gamma^D} \\
  & && (\refl_t)^\D\,\hat{\gamma^D} && :\equiv \hat{\refl}_{\hat{(}t^\D\,\hat{\gamma^D}\hat{)}} \\
  & && (\J_{a\,t\,(x.z.p)}\,pr\,_u\,eq)^\D\,\hat{\gamma^D} && :\equiv \hat{\J}\,\hat{\big(}\hat{\J}\,\hat{(}pr^\D\,\hat{\gamma^D}\hat{)}\,\hat{(}eq^\A\,\hat{\gamma}\hat{)}\hat{\big)}\,\hat{(}eq^\D\,\hat{\gamma^D}\hat{)} \\
  & && (\J\beta_{a\,t\,(x.z.p)}\,pr)^\D\,\hat{\gamma^D} && :\equiv \hat{\refl} \\
  & (5) && (a=_\U b)^\D\,\hat{\gamma^D}\,\hat{e} && :\equiv \TR_{\hat{(\lambda A.A\ra\Type_0)}}\,\hat{e}\,\hat{(}a^\D\,\hat{\gamma^D}\hat{)} \eqh b^\D\hat{\gamma^D} \\
  & && (\refl_a)^\D\,\hat{\gamma^D} && :\equiv \hat{\refl}_{\hat{(}a^\D\,\hat{\gamma^D}\hat{)}} \\
  & && (\J_{a\,(x.z.p)}\,pr\,_b\,eq)^\D\,\hat{\gamma^D} && :\equiv \hat{\J}\,\hat{\big(}\hat{\J}\,\hat{(}pr^\D\,\hat{\gamma^D}\hat{)}\,\hat{(}eq^\A\,\hat{\gamma}\hat{)}\hat{\big)}\,\hat{(}eq^\D\,\hat{\gamma^D}\hat{)} \\
  & && (\J\beta_{a\,t\,(x.z.p)}\,pr)^\D\,\hat{\gamma^D} && :\equiv \hat{\refl} \\
  & (6) && ((\hat{x}\in \hat{A})\ra B)^\D\,\hat{\gamma^D}\,\hat{f} && :\equiv \hat{(}\hat{x}\in \hat{A}\hat{)}\rah  B^\D\,\hat{\gamma^D}\,\hat{(}\hat{f}\,\hat{x}\hat{)} \\
  & && (t\,\hat{u})^\D\,\hat{\gamma^D} && :\equiv t^\D\,\hat{\gamma^D}\,\hat{u} \\
  & (7) && ((\hat{x}\in \hat{A})\ra b)^\D\,\hat{\gamma^D}\,\hat{f} && :\equiv \hat{(}\hat{x}\in \hat{A}\hat{)}\rah  b^D\,\hat{\gamma^D}\,\hat{(}\hat{f}\,\hat{x}\hat{)} \\
  & && (t\,\hat{u})^\D\,\hat{\gamma^D} && :\equiv t^\D\,\hat{\gamma^D}\,\hat{u}
\end{alignat*}
\endgroup
The predicate for a context is given by iterating $\blank^\D$ for its
constituent types. For a variable, the corresponding witness is looked up from
$\hat{\gamma^D}$.

The translation of the universe, given an element of
$\hat{A}\in\U^\A\,\hat{\gamma}$ (with
$\U^\A\,\hat{\gamma}\equiv\Type_{\hat{0}}$) returns the predicate space over
$\hat{A}$. For $\underline{a}$ types, we just return the translation of $a$.

The predicate for a function type for inductive parameters expresses
preservation of predicates. Witnesses of application are given by recursive
application of $\blank^\D$. The definitions for the other (non-inductive)
function spaces are similar, except there is no predicate for the domain types,
and thus no witnesses are required.

The translation for paths between $a : \U$ elements $t=_a u$, for each
$\hat{e}\in(t=_a u)^\A\,\hat{\gamma}$, i.e.\ $\hat{e}\in t^\A\,\hat{\gamma}\eqh
u^\A\,\hat{\gamma}$, witnesses that $t^\D$ and $u^\D$ are equal. As these have
different types, we have to transport over the original equality
$\hat{e}$. Hence, induction methods for path constructors will be \emph{paths
  over paths} in the sense of homotopy type theory.

$\refl$ is interpreted with just reflexivity in the target syntax. The
interpretation of $\J$ is given by a double $\hat{\J}$ application, borrowing
the definition from Lasson \cite{lasson}. Here, we use a shortened $\hat{\J}$
notation; see the formalization (Section \ref{sec:formalization}) for details.

For $a =_\U b$ paths, the interpretation is essentially the same as with the
other path type.

Again, let us consider the circle example:
\begin{alignat*}{5}
  & && (\boldsymbol{\cdot},\,S^1 : \U,\,b:\underline{S^1},\,loop: \underline{b=b})^\D\,\,\hat{(}\hat{\tt,\,S^1,\,b,\,loop}\hat{)}\\
  & \equiv \,\, && \hat{\top}\timesh\hat{(}\hat{S^{1D}}\in \hat{S^1}\rah\Type_{\hat{i}}\hat{)}\timesh\hat{(}\hat{b^D}\in \hat{S^{1D}\,b}\hat{)}\timesh\hat{(}\hat{loop^D\in \TR_{\,S^{1D}}\,loop\,b^D = b^D}\hat{)}
\end{alignat*}
The inputs of $\blank^\D$ here are the signature for the circle (the context in
black) and an $\hat{S^{1}}$-algebra consisting of three non-$\top$ components.
It returns a family over the type $\hat{S^1}$, an element of this family
$\hat{b^D}$ at index $\hat{b}$, and a path between $\hat{b^D}$ and
$\hat{b^D}$ which lies over $\hat{loop}$. This is the same as the usual
induction motives and methods for the circle, e.g.\ as described in \cite{HoTTbook}.


\section{From Logical Relations to Homomorphisms}
\label{sec:morphisms}

In this section, we specify the $\blank^\M$ translation, which computes
homomorphisms of algebras. We do so by first considering a logical relation
interpretation, and then refining it towards homomorphisms. For dependent type
theories, logical relation models are well-known (see e.g.\ \cite{atkey}), so
they are certainly applicable to our restricted syntax as well.

Below, we list induction motives for $\blank^\M$; these remain the same as we
move from logical relations to homomorphisms. The universe level $\hat{i}$
was chosen previously for the $\blank^\D$ translation.
\[
\infer{\hat{\Gamma}\vdashh\Delta^\M \in \Delta^\A\rah\Delta^\A\rah \Type_{\hat{i}}}
      {\hat{\Gamma}\vdash\Delta}
\hspace{1em}
\infer{
  \hat{\Gamma}\vdashh A^\M \in \hat{(}\hat{\gamma_0\,\gamma_1}\in\Delta^\A\hat{)}\rah \Delta^\M\,\hat{\gamma_0\,\gamma_1}\rah  A^\A\,\hat{\gamma_0}\rah A^\A\,\hat{\gamma_1}\rah \Type_{\hat{i}}}
      {\hat{\Gamma}\semicol\Delta\vdash A}
\]

\[
\infer{
  \hat{\Gamma}\vdashh t^\M\, \hat{ : (\gamma_0\,\gamma_1 : \blc{\Delta^\A})(\gamma^M : \blc{\Delta^\M}\,\gamma_0\,\gamma_1)\ra \blc{A^\M}\,\gamma_0\,\gamma_1\,\gamma^M\,(\blc{t^\A}\,\gamma_0)\,(\blc{t^\A}\,\gamma_1)
      }}
      {\hat{\Gamma}\semicol\Delta\vdash t : A}
\]

Contexts are mapped to (proof-relevant) relations and types to families of
relations depending on interpreted contexts. For terms, we again get a
fundamental theorem: every term has related standard interpretations in related
semantic contexts.

We present below the logical relation interpretation only for contexts, variables,
the universe and the inductive function space.

\begingroup
\allowdisplaybreaks
\begin{alignat*}{5}
  & (1)\hspace{1em} && \boldsymbol{\cdot}^\M\hat{\gamma_0\,\gamma_1} && :\equiv \hat{\top} \\
  & && (\Delta,x:A)^\M\,\hat{(\gamma_0,\alpha_0)\,(\gamma_1,\alpha_1)} && :\equiv \hat{(\gamma^M : \blc{\Delta^\M}\,\gamma_0\,\gamma_1)\times \blc{A^\M}\,\gamma_0\,\gamma_1\,\gamma^M\,\alpha_0\,\alpha_1} \\
  & && x^\M\,\hat{\gamma_0\,\gamma_1\,\gamma^M} && :\equiv x^{\text{th}}\text{ component in } \hat{\gamma^M} \\
  & (2) && \U^\M\,\hat{\gamma_0\,\gamma_1\,\gamma^M\,A_0\,A_1} && :\equiv \hat{A_0\ra A_1 \ra \Type_0} \\
  & && (\underline{a})^\M\,\hat{\gamma_0\,\gamma_1\,\gamma^M\,t_0\,\,t_1} && :\equiv \hat{
    \blc{a^\M}\,\gamma_0\,\gamma_1\,\gamma^M\,t_0\,\,t_1}\\
  & (3) && ((x:a)\ra B)^\M\,\hat{\gamma_0\,\gamma_1\,\gamma^M\,f_0\,f_1} && :\equiv
  \hat{(x_0 : \blc{(\underline{a})^\A}\,\gamma_0)(x_1 : \blc{(\underline{a})^\A}\,\gamma_1)(x^M : \blc{(\underline{a})^\M}\,\gamma_0\,\gamma_1\,\gamma^M)}\\
  & && && \hat{\hspace{1.7em}\ra \blc{B^\M}\,(\gamma_0,\,x_0)\,(\gamma_1,\,x_1)\,(\gamma^M,\,x^M)\,(f_0\,x_0)\,(f_1\,x_1)}
\end{alignat*}
\endgroup

We interpret the universe as relation space, and function types as relations
expressing pointwise relatedness of functions. This interpretation would work
the same way for unrestricted (non-strictly positive) function types as well.
For an example, this yields the following definition of logical relations
between natural number algebras:
\begin{alignat*}{5}
  & && (\boldsymbol{\cdot},Nat : \U,zero:\underline{Nat},suc:Nat\ra\underline{Nat})^\M
    \,\hat{(\tt,\,N_0,\,z_0,\,s_0)\,(\tt,\,N_1,\,z_1,\,s_1)} \\
    & \equiv\,\, && \hat{\top\times(N^M : N_0\ra N_1 \ra \Type_0)}\\
    & && \hat{\hspace{1em}\times\,\,(z^M : N^M\,z_0\,z_1)}\\
    & && \hat{\hspace{1em}\times\,\,(s^M : (x_0 : N_0)(x_1 : N_1)(x^M : N^M\,x_0\,x_1)\ra N^M\,(s_0\,x_0)\,(s_1\,x_1))}
\end{alignat*}

However, we would like to have underlying functions instead of relations in
homomorphisms. We take hint from the fact that for classical (simply-typed and
single-sorted) algebraic theories, a logical relation is equivalent to a
homomorphism if and only if the underlying relation is the graph of a function
\cite[pg. 5]{udayReynolds}. Thus we make the following change:
\begin{alignat*}{5}
  & \U^\M\,\hat{\gamma_0\,\gamma_1\,\gamma^M\,A_0\,A_1} && :\equiv \hat{A_0\ra A_1}
\end{alignat*}
This requires us to change $(\underline{a})^\M$ as well, since we need to
produce a type as result, but $a^M$ now yields a function. We can view the
result of $a^\M$ as a functional relation, and use its graph to relate $t_0$ and $t_1$:
\begin{alignat*}{5}
  & (\underline{a})^\M\,\hat{\gamma_0\,\gamma_1\,\gamma^M\,t_0\,\,t_1} && :\equiv \hat{(
    \blc{a^\M}\,\gamma_0\,\gamma_1\,\gamma^M\,t_0 = t_1)}
\end{alignat*}
At this point we have merely restricted relations to functions, and left the
rest of the interpretation unchanged. However, this is not strictly the desired
notion of homomorphism. Consider now again the $\blank^\M$ interpretation for natural
numbers:
\begin{alignat*}{5}
  & && (\boldsymbol{\cdot},Nat : \U,zero:\underline{Nat},suc:Nat\ra\underline{Nat})^\M
    \,\hat{(\tt,\,N_0,\,z_0,\,s_0)\,(\tt,\,N_1,\,z_1,\,s_1)} \\
    & \equiv\,\, && \hat{\top\times(N^M : N_0\ra N_1)}\\
    & && \hat{\hspace{1em}\times\,\,(z^M : N^M\,z_0 = z_1)}\\
    & && \hat{\hspace{1em}\times\,\,(s^M : (x_0 : N_0)(x_1 : N_1)(x^M : N^M\,x_0 = x_1)\ra N^M\,(s_0\,x_0)=s_1\,x_1)}
\end{alignat*}
In $\hat{s^M}$, there is a superfluous $\hat{x^M}$ equality proof. Fortunately,
in the translation of the inductive function space, we can just singleton
contract $\hat{x^M}$ away, yielding an equivalent, but stricter definition:
\begin{alignat*}{5}
  & ((x:a)\ra B)^\M\,\hat{\gamma_0\,\gamma_1\,\gamma^M\,f_0\,f_1} :\equiv \\
  & \hat{\hspace{2em}(x_0 : \blc{a^\A}\,\gamma_0)\ra \blc{B^\M}\,(\gamma_0,\,x_0)\,(\gamma_1,\,\blc{a^\M}\,\gamma_0\,\gamma_1\,\gamma^M\,x_0)\,(\gamma^M,\,\refl)\,
    (f_0\,x_0)\,(f_1\,(\blc{a^\M}\,\gamma_0\,\gamma_1\,\gamma^M\,x_0))}
\end{alignat*}
Now, the $\beta$-rule for successors is as expected:
\begin{alignat*}{5}
 & \hat{s^M : (x_0 : N_0)\ra N^M\,(s_0\,x_0)=s_1\,(N^M\,x_0)}
\end{alignat*}

Note that this singleton contraction is not possible for a general non-strictly
positive function space. We rely on the domain being small: $(\underline{a})^\M$
yields an equation, but for general $\hat{\Gamma;\,}\Delta\vdash A$ types,
$A^\M$ only yields an unknown relation.

$\blank^\M$ is similar to the translation to displayed algebra sections, which
is discussed in the next Section. We will discuss in more detail the
interpretations of equalities and the other (external) function types there. A
full listing for the homomorphism translation can be found in Appendix
\ref{sec:morphismrules}.


\section{Displayed Algebra Sections}
\label{sec:sections}

The operation $\blank^\S$ yields displayed algebra sections. Sections can be
viewed as dependent homomorphisms: while homomorphisms are structure-preserving
families of functions, sections are structure-preserving families of dependent
functions.

Contexts are interpreted as dependent relations between algebras and displayed
algebras. We again fix a universe level $\hat{i}$.
\[
\infer{\hat{\Gamma}\semicol\Delta^\S \in \hat{(}\hat{\gamma}\in\Delta^\A\hat{)}\rah \Delta^\D\,\hat{\gamma}\rah \Type_{\hat{i}}}{\hat{\Gamma}\vdash\Delta}
\]
Types are interpreted as dependent relations which additionally depend on
$\hat{\gamma}$, $\hat{\gamma^D}$, $\hat{\gamma^S}$ interpretations of the
context.
\[
\infer{\hat{\Gamma}\vdashh A^\S \in \hat{(}\hat{\gamma}\in\Delta^\A\hat{)}\hat{(}\hat{\gamma^D}\in\Delta^\D\,\hat{\gamma}\hat{)}\hat{(}\hat{\gamma^S}\in\Delta^\S\,\hat{\gamma}\,\hat{\gamma^D}\hat{)}\hat{(}\hat{x}\in A^\A\,\hat{\gamma}\hat{)}\rah  A^\D\,\hat{\gamma}\,\hat{\gamma^D}\,\hat{x}\rah \Type_{\hat{i}}}{\hat{\Gamma}\semicol\Delta\vdash A}
\]
For a term $t$, $t^\S$ again witnesses a fundamental theorem.
\[
\infer{\hat{\Gamma}\vdashh t^\S \in \hat{(}\hat{\gamma}\in\Delta^\A\hat{)}\hat{(}\hat{\gamma^D}\in\Delta^\D\,\hat{\gamma}\hat{)}\hat{(}\hat{\gamma^S}\in\Delta^\S\,\hat{\gamma}\,\hat{\gamma^D}\hat{)}\rah  A^\S\,\hat{\gamma}\,\hat{\gamma^D}\,\hat{\gamma^S}\,\hat{(}t^\A\,\hat{\gamma}\hat{)}\,\hat{(}t^\D\,\hat{\gamma}\,\hat{\gamma^D}\hat{)}}{\hat{\Gamma}\semicol\Delta \vdash t : A}
\]
We present the implementation below. Here, we make $\hat{\gamma}$-s and
$\hat{\gamma^D}$-s implicit and mostly notate $\hat{\gamma^S}$ parameters and
applications.

\begingroup
\allowdisplaybreaks
\begin{alignat*}{5}
  & (1)\hspace{1em} && \boldsymbol{\cdot}^\S\,\hat{\gamma}\,\hat{\gamma^D} && :\equiv \hat{\top} \\
  & && (\Delta,\,x:A)^\S\,(\hat{\gamma},\hat{t})\,(\hat{\gamma^D},\hat{t^D}) && :\equiv \hat{(}\hat{\gamma^S}\in\Delta^\S\hat{\gamma^2}\hat{)}\timesh A^\S\,\hat{\gamma^S}\,\hat{t}\,\hat{t^D} \\
  & && x^\S\,\hat{\gamma^S} && :\equiv x^{\text{th}}\text{ component in } \hat{\gamma^S} \\
  & (2) && \U^\S\,\hat{\gamma^S}\,\hat{A}\,\hat{A^D} && :\equiv \hat{(}\hat{x}\in \hat{A}\hat{)}\rah  \hat{A^D}\,\hat{x} \\
  & && (\underline{a})^\S\,\hat{\gamma^S}\,\hat{t}\,\hat{t^D} && :\equiv a^\S\,\hat{\gamma^S}\,\hat{t} \eqh \hat{t^D} \\
  & (3) && ((x:a)\ra B)^\S\,\hat{\gamma^S}\,\hat{f}\,\hat{f^D} && :\equiv \hat{(}\hat{x}\in a^\A\,\hat{\gamma}\hat{)}\rah  B^\S\,\hat{(}\hat{\gamma,x}\hat{)}\,\hat{(}\hat{\gamma^D,}\,a^\S\,\hat{\gamma^S}\,\hat{x}\hat{)}\,\hat{(}\hat{\gamma^S,}\,\hat{\refl}\hat{)} \\
  & && && \hspace{8.4em} \hat{(}\hat{f}\,\hat{x}\hat{)}\,\hat{\big(}\hat{f^D}\,\hat{x}\,\hat{(}a^\S\,\hat{\gamma^S}\,\hat{x}\hat{)}\hat{\big)} \\
  & && (t\,u)^\S\,\hat{\gamma^S} && :\equiv \hat{\J}\,\hat{(}t^\S\,\hat{\gamma^S}\,\hat{(} u^\A\,\hat{\gamma}\hat{)}\hat{)}\,\hat{(}u^\S\,\hat{\gamma^S}\hat{)} \\
  & (4) && (t=_a u)^\S\,\hat{\gamma^S}\,\hat{e} && :\equiv \TR\,\hat{(}t^\S\,\hat{\gamma^S}\hat{)}\hat{\big(}\TR\,\hat{(}u^\S\,\hat{\gamma^S}\hat{)}\,\hat{(}\apd\,\hat{(}a^\S\,\hat{\gamma^S}\hat{)}\,\hat{e}\hat{)}\hat{\big)} \\
  & && (\refl_t)^\S\,\hat{\gamma^S} && :\equiv \hat{\J}\,\hat{\refl}\,\hat{(}t^\S\,\hat{\gamma^S}\hat{)} \\
  & && (\J_{a\,t\,(x.z.p)}\,pr\,_u\,eq)^\S\,\hat{\gamma^S} && :\equiv \\
  & && && \hspace{-10em}\hat{\J}\,\hat{\bigg(}\hat{\J}\,\hat{\Big(}\hat{\J}\,\hat{\big(}\hat{\J}\,\hat{(}\hat{\lambda}\,\hat{p^D}\,\hat{p^S}\,\hat{pr^D}\,\hat{pr^S}\hat{.}\,\hat{pr^S}\hat{)}\,\hat{(}t^\S\,\hat{\gamma^S}\hat{)}\,\\
  & && && \hspace{-9em}\hat{(}\hat{\mathsf{uncurry}}\,p^\D\,\hat{\gamma^2}\hat{)}\,\hat{(}\hat{\mathsf{uncurry}}\,p^\S\,\hat{\gamma^S}\hat{)}\,\hat{(}pr^\D\,\hat{\gamma^2}\hat{)}\,\hat{(}pr^\S\,\hat{\gamma^S}\hat{)}\hat{\big)}\,\hat{(}eq^\A\,\hat{\gamma}\hat{)}\hat{\Big)}\,\hat{(}u^\S\,\hat{\gamma^S}\hat{)}\hat{\bigg)}\,\hat{(}eq^\S\,\hat{\gamma^S}\hat{)} \\
  & && (\J\beta_{a\,t\,(x.z.p)}\,pr)^\S\,\hat{\gamma^S} && :\equiv \\
  & && && \hspace{-10em}\hat{\J}\,\hat{\big(}\hat{\J}\,\hat{(}\hat{\lambda}\,\hat{p^D}\,\hat{p^S}\hat{.}\,\hat{\refl}\hat{)}\,\hat{(}t^\S\,\hat{\gamma^S}\hat{)}\,\hat{(}\hat{\mathsf{uncurry}}\,p^\D\,\hat{\gamma^2}\hat{)}\,\hat{(}\hat{\mathsf{uncurry}}\,p^\S\,\hat{\gamma^S}\hat{)}\hat{\big)}\,\hat{(}pr^\S\,\hat{\gamma^S}\hat{)}\\
  & (5) && (a=_\U b)^\S\,\hat{\gamma^S} && :\equiv
    \hat{\lambda\,e\,e^D.\,(\lambda\,x.\, \blc{b^\S}\,\gamma^S\, (\coe\,e\,x))
     =
    (\lambda\,x.\, \TR\,e^D\,(\J\,(\blc{a^\S}\,\gamma^S\,x)\,e))}\\
  & && (\refl_a)^\S\,\hat{\gamma^S} && :\equiv \hat{\refl} \\
  & && (\J_{a\,(x.z.p)}\,pr\,_b\,eq)^\S\,\hat{\gamma^S} && :\equiv\\
  & && && \hspace{-10em}\hat{\J\,\bigg(\lambda\,b^D\,\,b^S\,eq^D.\,\J\,\Big(\lambda\,b^S\,eq^S.\,\J\,(\lambda P^S\,pr^S.\,pr^S)\,eq^S\,(\mathsf{uncurry}\,\blc{p^\S}\,\gamma^S)\,(\blc{pr^\S}\,\gamma^S)\Big)}  \\
  & && && \hspace{-9em}\hat{\,(\blc{eq^\D}\,\gamma^D)\,b^S\bigg)\,(\blc{eq^\A}\,\gamma)\,(\blc{b^\D}\,\gamma^D)\,(\blc{b^\S}\,\gamma^S)\,(\blc{eq^\D}\,\gamma^D)\,(\blc{eq^\S}\,\gamma^S) } \\
  & && (\J\beta_{a\,t\,(x.z.p)}\,pr)^\S\,\hat{\gamma^S} && :\equiv \hat{\J\,\refl\,(\blc{pr^\S}\,\gamma^S)} \\
  & (6) && ((\hat{x}\in \hat{A})\ra B)^\S\,\hat{\gamma^S}\,\hat{f}\,\hat{f^D} && :\equiv \hat{(}\hat{x}\in \hat{A}\hat{)}\rah B^\S\,\hat{\gamma^S}\,\hat{(}\hat{f}\,\hat{x}\hat{)}\,\hat{(}\hat{f^D}\,\hat{x}\hat{)} \\
  & && (t\,\hat{u})^\S\,\hat{\gamma^S} && :\equiv t^\S\,\hat{\gamma^S}\,\hat{u} \\
  & (7) && ((\hat{x}\in \hat{A})\ra b)^\S\,\hat{\gamma^S\,f\,t} && :\equiv b^\S\,\hat{\gamma^S}\,\hat{(}\hat{f}\,\hat{t}\hat{)} \\
  & && (t\,\hat{u})^\S\,\hat{\gamma^S} && :\equiv \ap\,\hat{(}\hat{\lambda} \hat{f.}\hat{f}\,\hat{u}\hat{)}\,\hat{(}t^\S\,\hat{\gamma^S}\hat{)}
\end{alignat*}
\endgroup

The interpretations follow the same pattern as in the case of $\blank^\M$ up
until $((x:a)\ra B)^\S$, with $\U^\S$ defined as a dependent function instead of
a non-dependent one. Also, $\U^\S\,\hat{\gamma^S}\,\hat{A}\,\hat{A^D}$ is
precisely the type of sections of the $\hat{A^D}$ type family.

Let us consider now the translation in the
corresponding induction principles in mind, defined as $\mathsf{Induction}$ in
Section \ref{sec:general}.

The $\U^\S$ rule yields the type of the eliminator function for a type
constructor. For natural numbers, the non-indexed $Nat : \U$ is interpreted
as $\hat{\hat{Nat^S} \in (\hat{x}\in \hat{Nat})\rah \hat{Nat^D}\,\hat{x}}$. For
indexed types, the indices are first processed by the $\blank^\S$ cases for
inductive and external function parameters, until the ultimate $\U$ return type
is reached. Hence, we always get an eliminator function for a type constructor.

Analogously, the $\blank^\S$ result type for a point or path constructor is
always a $\beta$-rule, i.e.\ a function type returning an equality. That is
because $(\underline{a})^\S$ expresses that applications of $a^\S$ eliminators
must be equal to the corresponding $\hat{t^D}$ induction methods. Hence, for
path and point constructor types, $\blank^\S$ works by first processing all
inductive and external parameters, then finally returning an equality type.

In the case of $t =_a u$ equalities, we only provide abbreviated definitions for
the $t\,u$, $t=_a u$, $\refl$, $\J$ and $\J\beta$ cases. In the $\J$ case, we
write $\hat{\mathsf{uncurry}}\,p^\D$ for
$\hat{\lambda\,\gamma\,\gamma^D\,x\,x^D\,z\,z^D.}$ $\,p^\D\,\hat{(}\hat{\gamma,x,z}\hat{)}\,\hat{(}\hat{\gamma^D,x^D,z^D}\hat{)}$
and analogously elsewhere, to adjust for the fact that $p$ abstracts over
additional $x$ and $z$ variables. The full definitions can be found in the Agda
formalization. The definitions are highly constrained by the required types, and
not particularly difficult to implement with the help of a proof assistant: they
all involve doing successive path induction on all equalities available from
induction hypotheses, with appropriately generalized induction motives.

The full $(\J_{a\,t\,(x.z.p)}\,pr\,_u\,eq)^\S$ definition is quite large, and,
for instance, yields a very large $\beta$-rule for the higher inductive torus
definition (the reader can confirm this using the Haskell implementation). One
could have an implementation with specialized cases for commonly used operations
such as path compositions and inverses, in order to produce smaller translation
output.

The circle example is a bit more interesting here:
\begin{alignat*}{5}
  & && (\boldsymbol{\cdot},\,S^1 : \U,\,b:\underline{S^1},\,loop: \underline{b=b})^\S\,\,\hat{(}\hat{\tt,\,S^1,\,b,\,loop}\hat{)}\,\hat{(}\hat{\tt,\,S^{1D},\,b^D,\,loop^D}\hat{)} \\
  & \equiv \,\, && \hat{\top}\timesh\hat{(}\hat{S^{1S}}\in \hat{(}\hat{x}\in\hat{S^1}\hat{)}\hat{\,\ra\,}\hat{S^{1D}\,x}\hat{)}\timesh\hat{(}\hat{b^S}\in \hat{S^{1S}\,\,b = b^D}\hat{)}\\
  & && \hspace{0.75em}\timesh\hat{(}\hat{loop^S\,\,\in\,\,} \TR\,_{\hat{(}\hat{\lambda x. \TR\,_{S^{1D}}\,loop\,x\,=\,b^D}\hat{)}}\,\hat{b^S}\,\hat{(}\TR\,_{\hat{(}\hat{\lambda} \hat{x}. \TR\,_{\hat{S^{1D}}}\,\hat{loop}\,\hat{(}\hat{S^{1S}\,b}\hat{)}\,\eqh\,\hat{x}\hat{)}}\,\hat{b^S}\,\hat{(}\apd\,\hat{S^{1S}\,loop}\hat{)}\hat{)}\,\\
  & && \hspace{5.5em} \hat{=\,loop^D}\hat{)}
\end{alignat*}

In homotopy type theory, the $\beta$-rule for $loop$ is usually just
$\apd\,\hat{S^{1E}\,loop}\,\hat{\,=\,loop^D}$, but here all
$\beta$-rules are propositional, so we need to transport with
$\hat{b^S}$ to make the equation well-typed. When computing the type
of $\hat{loop^S}$, we start with $(\underline{b
  =b})^S\,\,\hat{\gamma^3}\,\,\hat{loop}\,\,\hat{loop^D}$. Next, this
evaluates to $(b=b)^\S\,\hat{\gamma^3}\,\hat{loop}\eqh\hat{loop^D}$,
and then we unfold the left hand side to get the doubly-transported
expression in the result.

For $a =_\U b$ equalities, let us examine here only the $(a =_\U b)^\S$ case.
The interpretation is inspired by the univalence axiom, although the translation
does not require it. If we assume univalence in the external theory, then
$\hat{a =_{\Type_i} b}$ equalities are equivalent to equivalences, which
contain $\hat{f : a \ra b}$ functions together with $\hat{\mathsf{isEqv}\,f}$
proofs. Hence, we can view the $\hat{e : \blc{a^\A}\,\gamma =
  \blc{b^\A}\,\gamma}$ and $\hat{e^D : \TR_{(\lambda\,A.\,A \ra
    \mathsf{Type_0})}\,e\,(\blc{a^\D}\,\gamma^D) = \blc{b^\D}\,\gamma^D}$ proofs
as functions bundled together with witnesses of equivalence. Since
$\hat{\mathsf{isEqv}}$ is proof irrelevant, we only need to relate the
underlying functions in the translation. For $\hat{e}$, the underlying function
is given just by $\hat{\coe\,e}$. For $\hat{e^D}$, the underlying function is a
bit more complicated, since $\hat{e^D}$ is a path over a path. We can observe
this in the example for higher inductive integers:
\begin{alignat*}{5}
  & && (\boldsymbol{\cdot},\,\,Int : U,\,\,zero : \underline{Int},\,p : Int =_\U Int)^\S\,
     \hat{(\tt,\,Int,\,z,\,p)\,(\tt,\,Int^D,\,z^D,\,p^D)} \\
  & \equiv \,\, && \hat{
     \top \times (Int^S : (x : Int)\ra Int^D\,x) \times (z^S : Int^S\,z = z^D)}\\
  & && \hat{\hspace{1em}\times\,
       (p^S : (\lambda\,x.\,Int^S\,(\coe\,p\,x)) = (\lambda\,x.\,\TR\,p^D\,(\J\,(Int^S\,x)\,p)))}
\end{alignat*}

In Appendix \ref{sec:app}, we additionally show how the type of displayed
algebra sections is computed for the two-dimensional sphere. In Appendix
\ref{sec:wtypes}, we show the same for indexed W-types.


\section{Possible Extensions to Categorical Semantics}
\label{sec:categorical}

So far, we were able to compute algebras, displayed algebras,
morphisms and sections, and this allows us to state recursion and
induction principles.  Reiterating Section \ref{sec:general}, for a
signature $\hat{\Gamma\vdash}\,\Delta$ and assuming $\hat{\Gamma\vdash
  \Delta^* : \blc{\Delta^\A}}$ as a candidate algebra for the HIIT, we
have the following types for induction and recursion:
\begin{alignat*}{5}
& \hat{\Gamma\vdash \mathsf{Induction} :
    (\gamma^D : \blc{\Delta^\D}\,\Delta^*)\ra \blc{\Delta^\S}\,\Delta^*\,\gamma^D}\\
& \hat{\Gamma\vdash \mathsf{Recursion} :
    (\gamma : \blc{\Delta^\A})\ra \blc{\Delta^\M}\,\Delta^*\,\gamma}
\end{alignat*}

However, this is not the full picture. We would also like to have a
\emph{category} of algebras, with homomorphisms as morphisms. This is not
without difficulties.

We are working in an UIP-free setting. In such setting, standard definitions of
categories feature set-truncated morphisms \cite{HoTTbook}. But we have
non-truncated notions of algebras and homomorphisms, so we cannot use the
standard definitions. One solution is to simply use non-truncated categories;
these have been previously called ``precategories'' or ``wild categories''
\cite{semisegal}. Sojakova demonstrated \cite{sojakova} that working with wild
categories internally to type theory is enough to prove equivalence of induction
and homotopy initiality for a class of higher inductive types, which suggests
that the same might be possible for HIITs. Still, it would be desirable to build
semantics of HIITs in a richer ($\omega$, 1)-categorical setting.

However, out current approach does not scale to the point where we have to worry
about higher categories. We explain in the following. The natural next step
towards a categorical semantics would be defining a $\blank^{\mathsf{ID}}$ translation,
which computes identity homomorphisms:

\[
\infer{\hat{\Gamma}\vdashh\Delta^{\mathsf{ID}}\,\hat{:\,(\gamma : \blc{\Delta^\A})\ra \blc{\Delta^\M}\,\gamma\,\gamma}}
      {\hat{\Gamma}\vdash\Delta}
\hspace{1em}
\infer{
  \hat{\Gamma}\vdashh A^{\mathsf{ID}}\,\hat{:\,(\gamma : \blc{\Delta^\A})(t : \blc{A^\A}\,\gamma)\ra \blc{A^\M}\,\gamma\,\gamma\,(\blc{\Delta^{\mathsf{ID}}}\,\gamma)\,t\,t }}
      {\hat{\Gamma}\semicol\Delta\vdash A}
\]

\[
\infer{
  \hat{\Gamma}\vdashh t^{\mathsf{ID}}\,\hat{:\, (\gamma : \blc{\Delta^\A})\ra
       \blc{t^\M}\,\gamma\,\gamma\,(\blc{\Delta^{\mathsf{ID}}}\,\gamma) = \blc{A^{\mathsf{ID}}}\,\gamma\,(\blc{t^\A}\,\gamma) }}
      {\hat{\Gamma}\semicol\Delta\vdash t : A}
\]

Here, the interpretation of terms witnesses functoriality: $t^\M$ maps
identity morphisms in $\Delta$ to displayed identity morphisms in $A$.
We translate $\U$ to identity functions:
\[
\U^{\mathsf{ID}}\,\hat{\gamma\,A} :\equiv \hat{\lambda\,(x : A).\, x}
\]
Assume that $\sigma$ is a parallel substitution, and note that $\U$ and $\U[\sigma]$
are definitionally equal in the theory of signatures. The translation of the
latter would be the following (omitting many details, including the handling of
substitutions in the translation):
\[
(\U[\sigma])^{\mathsf{ID}}\,\hat{\gamma\,A} \equiv \hat{\TR_{(\lambda\,x. A\ra A)}\,(\blc{\sigma^{\mathsf{ID}}}\,\gamma)\,(\lambda\,x.\, x)}
\]
Here, $\hat{\sigma^{\mathsf{ID}}\,\gamma}$ yields an equation for functoriality of
$\hat{\sigma}$. Since the transport is constant, the result is propositionally
equal to $\hat{\lambda\,x.\, x}$. However, that is not enough, since we need to
preserve $\U = \U[\sigma]$ up to definitional equality.

Alternatively, we could try changing the $\blank^\mathsf{ID}$ interpretation of
terms and substitutions to get \emph{definitional} equalities, instead of
propositional equalities internal to the target syntax. This would not be a
purely syntactic translation anymore, since for each
$\hat{\Gamma}\semicol\Delta\vdash t : A$ we would get a universally quantified
metatheoretic statement expressing that for each $\hat{\gamma}$ term,
$\hat{\blc{t^\M}\,\gamma\,\gamma\,(\blc{\Delta^{\mathsf{ID}}}\,\gamma)}$ is
definitionally equal to
$\hat{\blc{A^{\mathsf{ID}}}\,\gamma\,(\blc{t^\A}\,\gamma)}$. This would solve
the strictness problem in the case of $\U[\sigma]$, since there would be no need
to transport over definitional equalities.

Unfortunately, while this approach repairs $\blank^{\mathsf{ID}}$ in some cases,
other cases become unfeasible, because we are not able to produce definitional
equations. For example, the interpretation of the identity type would be as follows:
\[
(t =_a u)^{\mathsf{ID}}\,\hat{\gamma \,\,\blc{:}\,\, \blc{(t =_a u)^\M}\,\gamma\,\gamma\,(\blc{\Gamma^{\mathsf{ID}}}\,\gamma)\,(\blc{(t =_a u)^\A}\,\gamma)} \equiv \hat{\blc{\U^{\mathsf{ID}}}\,\gamma\,(\blc{(t =_a u)^\A}\,\gamma)}
\]
The goal type can be reduced along definitions, and using $t^{\mathsf{ID}}$ and $u^{\mathsf{ID}}$, to
\[
\hat{(\lambda\,e.\,\,\refl\,\sqcdot\,\ap\,(\lambda\,x.\,x)\,e\,\sqcdot\,\refl)\,\blc{\equiv}\,(\lambda\,e.\,e)}
\]
This definitional equality is clearly not provable, although $(t =_a
u)^{\mathsf{ID}}$ does work when we interpret terms with weak internal
equalities. Hence, there are strictness problems both with weak and strict
equations for functoriality.

We consider three potential solutions:
\begin{enumerate}
\item Solving the coherence problem. This would allow us to interpret signatures
      into the metatheory, allowing preservation of definitional equality up
      to propositional metatheoretic equality.
\item Reformulating the syntax of signatures so that definitional equalities
      become weak propositional equalities. We already use weak $\beta$ for $\J$,
      can we do so elsewhere? However, this would result in an unusual and very
      inconvenient syntax of signatures, because we would need to weaken even basic
      substitution rules. In contrast, weak $\beta$ for $\J$ seems harmless, because
      we are not aware of any HIIT signature in the literature that involves
      any $\J$ computation.
\item Instead of interpreting signatures into an UIP-free type theory, we
      interpret them into classical combinatorial structures for higher groupoids
      and categories, e.g.\ into simplicial sets. The drawback is that now we do not
      have the convenient synthetic notion of higher structures, provided by the syntax
      of type theory, and instead we have to manually build up these
      structures. Needless to say, this makes machine-checked formalization much
      more difficult. We have found mechanized formalization invaluable for the
      current paper, and it would be painful to abandon it in further research.
\end{enumerate}

\noindent
In summary, solutions for the strictness problems require significant deviation
from the approach of the current paper, or require significant further research.


\section{Formalization and Implementation}
\label{sec:formalization}

There are additional development artifacts to the current work: a Haskell
implementation, two Agda formalizations of the syntactic translations (a shallow
and a deeper version) and an Agda formalization for deriving a Frobenius $\J$
rule from $\J$. All are available from
\url{https://github.com/akaposi/hiit-signatures}.

The Haskell implementation takes as input a file which contains a of a
$\hat{\Gamma}\vdash\Delta$ signature. Then, it checks the input with respect to
the rules in Figures \ref{sigrules1} and \ref{sigrules2}, and outputs an
Agda-checkable file which contains algebras, homomorphisms, displayed algebras
and sections for the input signature.

It comes with examples, including the ones in this paper, the
inductive-inductive dense completion \cite[Appendix A.1.3]{forsberg-phd} and
several HITs from \cite{HoTTbook} including the definition of Cauchy reals. It
can be checked that our implementation computes the expected elimination
principles in these cases.

The shallow Agda formalization embeds both the source and target theories
shallowly into Agda: it represents types as Agda types, functions as Agda
functions, and so on. We also leave the $\blank^\A$ operation implicit. We state
each case of translations as Agda functions from all induction hypotheses to the
result type of the translation, which lets us ``typecheck'' the translation. We
have found that this style of formalization is conveniently light, but remains
detailed enough to be useful. We also generated some of the code of the Haskell
implementation from this formalization.

The deep Agda formalization deeply embeds the theory of signatures as a CwF with
additional structure, in the style of \cite{ttintt}. However, it embeds the
external type theory shallowly as Agda, and we model dependency on external
$\hat{\Gamma}$ contexts with Agda functions. We formalize $\blank^\A$,
$\blank^\D$, $\blank^\M$ and $\blank^\S$ translations as a single model
construction in Agda. This setup greatly simplifies formalization, since we do
not have to reason explicitly about definitional equality in the external
syntax, but we can still reason directly about preservation of definitional
equalities in the theory of signatures. This semi-deep formalization can be in
principle converted into a fully formal syntactic translation, because we prove
all preservations of definitional equalities by $\refl$. Due to technical issues
and performance issues in Agda, this deeper formalization uses transport instead
of $\J$ for $t =_a u$ equalities, and it does not cover elimination for $a
=_\U b$ equalities.


\section{Conclusions}
\label{sec:summary}

Higher inductive-inductive types are useful in defining the well-typed syntax of
type theory in an abstract way \cite{ttintt}. From a universal algebraic point of
view, they provide initial algebras for multi-sorted algebraic theories where
the sorts can depend on each other. From the perspective of homotopy type
theory, they provide synthetic versions of homotopy-theoretic constructions such
as higher-dimensional spheres or cell complexes. So far, no general scheme of
HIITs have been proposed. To quote Lumsdaine and Shulman
\cite{lumsdaineShulman}:
\begin{quotation}
``The constructors of an ordinary inductive type are unrelated to each
other.  But in a higher inductive type each constructor must be able
to refer to the previous ones; specifically, the source and target of
a path-constructor generally involve the previous
point-constructors. No syntactic scheme has yet been proposed for this
that covers all cases of interest while remaining meaningful and
consistent.''
\end{quotation}
In this paper we proposed such a syntactic scheme which also includes
inductive-inductive types. We tackled the problem of complex dependencies on
previous type formation rules and constructors by a well-known method of
describing intricate dependencies: the syntax of type theory itself. We had to
limit the type formers to only allow strictly positive definitions, but these
restrictions are the only things that a type theorist has to learn to understand
our signatures. Our encoding is also direct in the sense that notions of
induction and recursion are computed exactly as required and not merely up to
equivalences or isomorphisms, and we also demonstrated that this computation is
feasible to implement in computer programs.

We developed an approach where syntactic translations are used to provide
semantics for HIIT signatures. Our approach seems to be a sweet spot for
computing notions of induction and recursion in a formally verifiable and
relatively simple way.

However, there is a coherence problem in the formal treatment of syntaxes and
models of type theories internally to type theory. We sidestepped this by
considering syntactic translations, but the problem remains, and it prevents
extending the current approach to categorical semantics. Our impression is that
the true solution will be the development of higher syntaxes and models of type
theory in type theory. This may also require adding new features to the
meta type theory. In particular, convenient formalization of higher categories
seems elusive in conventional homotopy type theory, and we may need two-level
type theories \cite{semisegal}, or directed type-theories with native
notions of higher categories \cite{nuyts2015towards, riehl2017type}.

\bibliographystyle{plain}
\bibliography{b}

\appendix
\section{The homomorphism translation}
\label{sec:morphismrules}

The specification of the $\blank^\M$ interpretation is as following.
\[
\infer{\hat{\Gamma}\vdashh\Delta^\M \in \Delta^\A\rah\Delta^\A\rah \Type_{\hat{i}}}
      {\hat{\Gamma}\vdash\Delta}
\hspace{1em}
\infer{
  \hat{\Gamma}\vdashh A^\M \in \hat{(}\hat{\gamma_0\,\gamma_1}\in\Delta^\A\hat{)}\rah \Delta^\M\,\hat{\gamma_0\,\gamma_1}\rah  A^\A\,\hat{\gamma_0}\rah A^\A\,\hat{\gamma_1}\rah \Type_{\hat{i}}}
      {\hat{\Gamma}\semicol\Delta\vdash A}
\]

\[
\infer{
  \hat{\Gamma}\vdashh t^\M\, \hat{ : (\gamma_0\,\gamma_1 : \blc{\Delta^\A})(\gamma^M : \blc{\Delta^\M}\,\gamma_0\,\gamma_1)\ra \blc{A^\M}\,\gamma_0\,\gamma_1\,\gamma^M\,(\blc{t^\A}\,\gamma_0)\,(\blc{t^\A}\,\gamma_1)
      }}
      {\hat{\Gamma}\semicol\Delta\vdash t : A}
\]

We list below the implementation of $\blank^\M$ on the full syntax. For brevity, we
mostly leave $\hat{\gamma_0}$ and $\hat{\gamma_1}$ implicit, notating only
$\hat{\gamma^M}$ where necessary.  We also omit some subscript parameters from
$\hat{\J}$ applications. The interpretation follows the same pattern as in the
case of $\blank^\S$; the main difference is the interpretation of the universe
as non-dependent function space. Additionally, the equality type former and
reflexivity admit slightly nicer translations, since the lack of type dependency
allows us to use composition and $\hat{\inv}$ (which was noted in Section
\ref{sec:external}) instead of raw transports and path induction.

\begingroup
\allowdisplaybreaks
\begin{alignat*}{5}
  & (1)\hspace{1em} && \boldsymbol{\cdot}^\M\hat{\gamma_0\,\gamma_1} && :\equiv \hat{\top} \\
  & && (\Delta,x:A)^\M\,\hat{(\gamma_0,\alpha_0)\,(\gamma_1,\alpha_1)} && :\equiv \hat{(\gamma^M : \blc{\Delta^\M}\,\gamma_0\,\gamma_1)\times \blc{A^\M}\,\gamma^M\,\alpha_0\,\alpha_1} \\
  & && x^\M\,\hat{\gamma^M} && :\equiv x^{\text{th}}\text{ component in } \hat{\gamma^M} \\
  & (2) && \U^\M\,\hat{\gamma^M\,A_0\,A_1} && :\equiv \hat{A_0\ra A_1} \\
  & && (\underline{a})^\M\,\hat{\gamma^M\,t_0\,\,t_1} && :\equiv \hat{
    \blc{a^\M}\,\gamma^M\,t_0 = t_1} \\
  & (3) && ((x:a)\ra B)^\M\,\hat{\gamma^M\,f_0\,f_1} && :\equiv \\
  & && && \hat{\hspace{-8em}(x_0 : \blc{a^\A}\,\gamma_0)\ra \blc{B^\M}\,(\gamma_0,\,x_0)\,(\gamma_1,\,\blc{a^\M}\,\gamma^M\,x_0)\,(\gamma^M,\,\refl)\,
    (f_0\,x_0)\,(f_1\,(\blc{a^\M}\,\gamma^M\,x_0))}\\
  & && (t\,u)^\M\,\hat{\gamma^M} && :\equiv  \hat{\J\,(\blc{t^\M}\,\gamma^M\,(\blc{u^\A}\,\gamma_0))\,(\blc{u^\M}\,\gamma^\M) }\\
  & (4) && (t=_a u)^\M\,\hat{\gamma^M} && :\equiv
    \hat{\lambda\,e.\,(\blc{t^\M}\,\gamma^M)^{-1}\sqcdot\ap\,(\blc{a^\M}\,\gamma^M)\,e\sqcdot\,\blc{u^\M}\,\gamma^M}\\
  & && (\refl_t)^\M\,\hat{\gamma^M} && :\equiv \hat{\inv\,(\blc{t^\M}\,\gamma^M)}\\
  & && (\J_{a\,t\,(x.z.p)}\,pr\,_u\,eq)^\M\,\hat{\gamma^M} && :\equiv\\
  & && && \hspace{-10em}\hat{\J\,\bigg(\J\,\Big(\J\,\big(\J\,(\lambda\,p_1\,p^M\,pr_1\,pr^M.\, pr^M)\,(\blc{t^\M}\,\gamma^M)}  \\
  & && && \hspace{-9em}\hat{
      (\mathsf{uncurry}\,\blc{p^\A}\,\gamma_1)\,(\mathsf{uncurry}\,\blc{p^\M}\,\gamma^M)\,
      (\blc{pr^\A}\,\gamma_1)\,(\blc{pr^\M}\,\gamma^M)\big)\,
        (\blc{eq^\A}\,\gamma_0)\Big)(\blc{u^\M}\,\gamma^M)\bigg)(\blc{eq^\M}\,\gamma^M)   }\\
  & && (\J\beta_{a\,t\,(x.z.p)}\,pr)^\M\,\hat{\gamma^M} && :\equiv\\
  & && && \hspace{-9em}\hat{
    \J\,\big(\J\,(\lambda\,p_1\,p^M.\,\refl)\,(\blc{t^\M}\,\gamma^M)\,
      (\mathsf{uncurry}\,\blc{p^\A}\,\gamma_1)\,
      (\mathsf{uncurry}\,\blc{p^\M}\,\gamma^M)\big)
      (\blc{pr^\M}\,\gamma^M)
    }\\
  & (5) && (a=_\U b)^\M\,\hat{\gamma^M} && :\equiv
    \hat{\lambda\,e_0\,e_1.\,(\lambda\,x.\,\blc{b^\M}\,\gamma^M\,(\coe\,e_0\,x)) = (\lambda\,x.\,\coe\,e_1\,(\blc{a^\M}\,\gamma^M\,x))}\\
  & && (\refl_a)^\M\,\hat{\gamma^M} && :\equiv \hat{\refl} \\
  & && (\J_{a\,(x.z.p)}\,pr\,_b\,eq)^\M\,\hat{\gamma^M} && :\equiv\\
  & && && \hspace{-10em}\hat{\J\,\bigg(\J\,\Big(\lambda\,b^M\,eq^M.\,\J\,(\lambda\,P^M\,pr^M.\,pr^M)\,eq^M\,(\mathsf{uncurry}\,\blc{p^\M}\,\gamma^M)\,(\blc{pr^\M}\,\gamma^M)\Big)}  \\
  & && && \hspace{-9em}\hat{\,(\blc{eq^\A}\,\gamma_1)\bigg)\,(\blc{eq^\A}\,\gamma_0)\,(\blc{b^\M}\,\gamma^M)\,(\blc{eq^\M}\,\gamma^M) } \\
  & && (\J\beta_{a\,t\,(x.z.p)}\,pr)^\M\,\hat{\gamma^M} && :\equiv \hat{\inv\,(\blc{pr^\M}\,\gamma^M)} \\
  & (6) && ((\hat{x:A})\ra B)^\M\,\hat{\gamma^M\,f_0\,f_1} && :\equiv \hat{(x : A)\ra \blc{B^\M}\,\gamma^M\,(f_0\,x)\,(f_1\,x)} \\
  & && (t\,\hat{u})^\M\,\hat{\gamma^M} && :\equiv t^\M\,\hat{\gamma^M\,u} \\
  & (7) && ((\hat{x}\in \hat{A})\ra b)^\M\,\hat{\gamma^M} && :\equiv
       \hat{\lambda f\,x.\,\,\blc{b^\M}\,\gamma^M\,(f\,x)} \\
  & && (t\,\hat{u})^\M\,\hat{\gamma^M} && :\equiv
       \hat{\ap\,(\lambda f.\,f\,u)\,(\blc{t^\M}\,\gamma^M)}
\end{alignat*}
\endgroup

\section{Displayed algebra sections for the two-dimensional sphere}
\label{sec:app}

The two-dimensional sphere is given by the following signature:

\[
\Gamma :\equiv \Big(\boldsymbol{\cdot},\,\,\,S^2:\U,\,\,\,b:\underline{S^2},\,\,\,surf:\underline{\refl_{b} =_{(b=_{S^2} b)} \refl_{b}}\Big)
\]
The sphere-algebras are computed as follows.
\begin{alignat*}{5}
  & && \Gamma^\A \equiv \hat{\top}\timesh\hat{(}\hat{S^2}\in\Type_{\hat{0}}\hat{)}\timesh\hat{(}\hat{b}\in \hat{S^2}\hat{)}\timesh\hat{(}\hat{surf\in \hat{\refl_{\hat{b}}}=_{\hat{(}\hat{b}=_{\hat{S^2}}\hat{b}\hat{)}}\hat{\refl_{\hat{b}}}}\hat{)}
\end{alignat*}
Given a sphere-algebra and fixing a universe level $\hat{i}$, the displayed algebra is the following.
\begin{alignat*}{5}
  & && \Gamma^\D\,\,\hat{(}\hat{\tt,\,S^2,\,b,\,surf}\hat{)}\\
  & \equiv \,\, && \hat{\top}\timesh\hat{(}\hat{S^{2D}}\in \hat{S^2}\rah\Type_{\hat{i}}\hat{)} \\
  & && \hspace{0.75em} \timesh\hat{(}\hat{b^D}\in \hat{S^{2D}\,b}\hat{)} \\
  & && \hspace{0.75em} \timesh\hat{(}\hat{surf^D\in \TR_{\hat{(}\TR_{S^{2D}}\,\blank\,b^D=b^D\hat{)}}\,surf\,\refl_{b^D} = \refl_{b^D}}\hat{)}
\end{alignat*}
Given a sphere-algebra and a displayed algebra over it, we get the type of sections:
\begin{alignat*}{5}
  & && \Gamma^\S\,\,\hat{(}\hat{\tt,\,S^2,\,b,\,surf}\hat{)}\,\hat{(}\hat{\tt,\,S^{2D},\,b^D,\,surf^D}\hat{)} \\
  & \equiv \,\, && \hat{\top}\timesh\hat{(}\hat{S^{2S}}\in \hat{(}\hat{x}\in\hat{S^2}\hat{)}\hat{\,\ra\,}\hat{S^{2D}\,x}\hat{)} \\
  & && \hspace{0.75em} \timesh\hat{(}\hat{b^S}\in \hat{S^{2S}\,\,b = b^D}\hat{)}\\
  & && \hspace{0.75em} \timesh\hat{\bigg(}\hat{surf^S\,\,\in\,\,} \TR\,\hat{\Big(}\hat{\J\,\refl\,b^S}\hat{\Big)}\,\hat{\Big(}\TR\,\hat{\big(}\hat{\J\,\refl\,b^S}\hat{\big)}\,\hat{\big(}\hat{\apd}\,\hat{(}\hat{\lambda} \hat{x}\hat{.}\TR\,\hat{b^S}\,\hat{(}\TR\,\hat{b^S}\,\hat{(}\hat{\apd}\,\hat{S^{2S}}\,\hat{x}\hat{)}\hat{)}\hat{)}\,\hat{surf}\hat{\big)}\hat{\Big)} \\
  & && \hspace{5.5em} \hat{=} \,\hat{surf^D}\hat{\bigg)}
\end{alignat*}
Note that if $\hat{b^S}$ is a definitional equality (that is, we have
$\hat{S^{2S}\,\,b}\,\hat{\,\equiv\,}\,\hat{b^D}$), the occurrences of $\hat{b^S}$
in the type of $\hat{surf^S}$ can be replaced by $\hat{\refl}$. In
this case the type of $\hat{surf^S}$ becomes the more usual
$\hat{\apd}\,\hat{(}\hat{\apd}\,\hat{S^{2S}}\hat{)}\,\hat{surf}\,\hat{=}\,\hat{surf^D}$.

\section{Displayed algebra sections for indexed W-types}
\label{sec:wtypes}

Indexed W-types can describe a large class of inductive definitions
\cite{morris09indexed}. Suppose we have the external context $\hat{I}
\in \Type_{\hat{0}}\commah\hat{S} \in \Type_{\hat{0}}\commah\hat{P} \in \hat{S} \rah
\Type_{\hat{0}}\commah\hat{out} \in \hat{S} \rah \hat{I}\commah\hat{in} \in
\hat{(}\hat{s} \in \hat{S}\hat{)}\rah \hat{P}\,\hat{s}\rah \hat{I}$. Then, the
signature for the corresponding indexed W-type is the following:
\[
W :\equiv (\boldsymbol{\cdot},\,\,\,w:(\hat{i} \in \hat{I})\ra\U,\,\,\,sup: (\hat{s} \in \hat{S})\ra((\hat{p} \in \hat{P\,s})\ra w\,(\hat{in\,s\,p}))\ra \underline{w\,(\hat{out\,s})})
\]
We pick a universe level $j$ for elimination. The $\blank^\A$, $\blank^\D$ and $\blank^\S$ interpretations of $W$ are the following, omitting leading $\hat{\top}$ components:
\begin{alignat*}{5}
  & \rlap{$W^\A \equiv \hat{(}\hat{w}\in \hat{I}\rah\Type_{\hat{0}}\hat{)}\timesh\hat{(}\hat{(}\hat{s}\in \hat{S}\hat{)}\rah\hat{(}\hat{(}\hat{p}\in \hat{P\,s}\hat{)}\rah \hat{w}\,\hat{(}\hat{in\,s\,p}\hat{)}\hat{)}\rah \hat{w}\,\hat{(}\hat{out\,s}\hat{)}\hat{)}$}\\
  & W^\D\,\hat{(}\hat{w, sup}\hat{)} && \equiv\, && \hat{(}\hat{w^D} \in \hat{(}\hat{i}\in \hat{I}\hat{)}\rah \hat{w\,i}\rah \Type_{\hat{j}}\hat{)}\\
  & && \hspace{0.23em}\timesh && \hat{\big(}\hat{(}\hat{s\in S}\hat{)}\hat{(}\hat{f}\in \hat{(}\hat{p\in P\,s}\hat{)}\rah \hat{w}\,\hat{(}\hat{in\,s\,p}\hat{)}\hat{)}\\
  & && && \rah \hat{(}\hat{(}\hat{p\in P\,s}\hat{)}\rah \hat{w^D}\,\hat{(}\hat{in\,s\,p}\hat{)}\,\hat{(}\hat{f\,p}\hat{)}\hat{)}\rah \hat{w^D}\,\hat{(}\hat{out\,s}\hat{)}\,\hat{(}\hat{sup\,s\,f}\hat{)}\hat{\big)} \\
  & \rlap{$ W^\S\,\hat{(}\hat{w, sup}\hat{)}\,\hat{(}\hat{w^D,sup^D}\hat{)} \equiv $} \\
  & && && \hat{(}\hat{w^S} \in \hat{(}\hat{i \in I}\hat{)}\hat{(}\hat{x \in w\,i}\hat{)}\rah \hat{w^D\,i\,x}\hat{)}\\
  & && \hspace{0.23em}\timesh && \hat{\big(}\hat{(}\hat{s\in S}\hat{)}\hat{(}\hat{f}\in \hat{(}\hat{p\in P\,s}\hat{)}\rah \hat{w}\,\hat{(}\hat{in\,s\,p}\hat{)}\hat{)}\\
  & && && \rah \hat{w^S}\,\hat{(}\hat{out\,s}\hat{)}\,\hat{(}\hat{sup\,\,s\,f}\hat{)} \eqh \hat{sup^D\,s\,f}\,\hat{(}\hat{\lambda} \hat{p.}\, \hat{w^S}\,\hat{(}\hat{in\,s\,p}\hat{)}\,\hat{(}\hat{f\,p}\hat{)}\hat{)}\hat{\big)}
\end{alignat*}

\end{document}